\documentclass[prb,twocolumn,superscriptaddress, preprintnumbers]{revtex4}

\usepackage{times}
\usepackage{amsmath}
\usepackage{amsthm}
\usepackage{amssymb}
\usepackage{amsbsy}
\usepackage{amsfonts}
\usepackage{graphicx}

\begin{document}


\title{Short-Range Entangled Bosonic States with Chiral
Edge Modes and $T$-duality of Heterotic Strings}

\author{Eugeniu Plamadeala}
\affiliation{Department of Physics, University of California, Santa Barbara,
California 93106, USA}
\author{Michael Mulligan}
\affiliation{Microsoft Research, Station Q, Elings Hall,
University of California, Santa Barbara, California 93106-6105, USA}
\author{Chetan Nayak}
\affiliation{Microsoft Research, Station Q, Elings Hall,
University of California, Santa Barbara, California 93106-6105, USA}
\affiliation{Department of Physics, University of California, Santa Barbara, California 93106, USA}

\begin{abstract}
We consider states of bosons in two dimensions that do not support anyons in the bulk, but nevertheless have stable chiral edge modes that are protected even without any symmetry.
Such states must have edge modes with central charge $c=8k$ for integer $k$.
While there is a single such state with $c=8$, there are, naively,
two such states with $c=16$, corresponding to the
two distinct even unimodular lattices in $16$ dimensions. 
However, we show that these two phases are the same in the bulk,
which is a consequence of the uniqueness of signature $(8k +n, n)$ even unimodular lattices.
The bulk phases are stably equivalent, in a sense that we make precise.
However, there are two different phases of the edge corresponding
to these two lattices, thereby realizing a novel form of the bulk-edge
correspondence. Two distinct fully chiral edge phases are associated
with the same bulk phase, which is consistent with the uniqueness of the bulk
since the transition between them, which is generically first-order,
can occur purely at the edge.
Our construction is closely related to
$T$-duality of toroidally compactified heterotic strings.
We discuss generalizations of these results.
\end{abstract}

\maketitle

\section{Introduction}

The last decade has seen enormous progress in the understanding
of topological phases (see Ref. \onlinecite{Nayak08} and references therein) and of symmetry-protected topological (SPT) phases \cite{Chen11a,Chen11b,Kitaev11,Lu12}.
SPT phases are gapped phases of matter that do not have non-trivial excitations in the bulk; 
have vanishing topological entanglement entropy \cite{kitaevpreskill, levinwenentanglement} or, equivalently, have short-ranged entanglement (SRE); but have gapless excitations at the edge in the presence of a symmetry.
In the case of the most famous and best-understood example, `topological insulators' (see Refs. \onlinecite{kanemelez2, kanemelegraphene, bhz2006, fukanemele2007, moorebalents2007, Roy2009, qhz2008, ludwigclass, Hasan10,Qi11} and references therein), the symmetry is time-reversal. 
Topological phases (without a modifier) are gapped phases of matter
that are stable to arbitrary perturbations; support anyons in the bulk;
and have non-zero topological entanglement entropy
or, equivalently, have long-ranged entanglement (LRE).
They may or may not (depending on the topological phase)
have gapless edge excitations.\footnote{We note that SPT phases
can all be adiabatically connected to a trivial ground state if we
do not require that the associated symmetry be preserved. Topological
phases cannot be. However, if we restrict to Hamiltonians that respect
a symmetry then, just as the trivial phase splits into many SPT phases,
a non-trivial topological phase could split into multiple phases that could
be distinguished, for instance, by their edge excitations. For a discussion
of such ``symmetry-enhanced topological phases'', see Ref \onlinecite{Lu13}.}

However, there is a third possibility: phases of matter that do not support anyons
but nevertheless have gapless excitations even in the absence of
any symmetry. Thus, they lie somewhere between topological phases
and symmetry-protected topological phases, but are neither. Integer quantum
Hall states of fermions are a well-known example.
Their gapless edge excitations \cite{Laughlin81, Halperin82}
are stable to arbitrary weak perturbations even though they do not support anyons
and only have SRE\footnote{Note, however, that according to an alternate definition
of SRE states -- adiabatic continuability to a local product state
with finite-depth local unitary transformations\cite{Chen11a} -- integer
quantum Hall states of fermions and the bosonic state discussed in this paper would
be classified as LRE states.}.
Although the existence and stability of SRE
integer quantum Hall states might seem to be a special feature of fermions,
such states also exist in purely bosonic systems, albeit with some peculiar features.

For any integer $N$, there is an integer quantum Hall state of fermions
with SRE, electrical Hall conductance $\sigma_{xy} = N\frac{e^2}{h}$,
and thermal Hall conductance $\kappa_{xy}=N\frac{\pi^2 k_B^2 T} {3 h}$. \cite{kanefisher97}
In fact, there is only one such state for each $N$:
any two SRE states of fermions at the same filling fraction $N$
can be transformed into each other without encountering a phase transition.
\footnote{Of course, it may be possible to take a route from one to the other
that does cross a phase transition but such a transition can always be avoided.
For instance, if we restrict to $S_z$-conserving Hamiltonians,
then a phase transition must be encountered in going from
a spin-singlet $N=2$ state to a spin-polarized one. If we do not make this
restriction, however, then this phase transition can be avoided
and the two states can be adiabatically-connected.}
(This is true in the bulk; see Section \ref{sec:future} for the situation at the edge.)
Therefore, the state with $N$ filled Landau levels of non-interacting fermions
is representative of an entire universality class of SRE states.
As a result of its $N$ chiral Dirac fermion edge modes,
this is a distinct universality class from ordinary band insulators.
These edge modes, which have Virasoro central charge $c=N$ if all of the velocities are equal, are stable to {\it all}
perturbations. If we do not require charge conservation symmetry,
then some Hamiltonians in this universality class may not
have $\sigma_{xy}=N\frac{e^2}{h}$, but they will all have
$\kappa_{xy}=c\frac{\pi^2 k_B^2 T} {3 h}=N\frac{\pi^2 k_B^2 T} {3 h}$.

Turning now to bosons, there are SRE states of bosons with
similarly stable chiral edge modes, but only for central charges $c=8k$.
As we discuss, they correspond to even, positive-definite, unimodular lattices.
Moreover, while there is a unique such state with $c=8$, there
appear to be two with $c=16$, twenty-four with $c=24$, and more than ten million
with $c=32$. \cite{conwaysloane} Thus, we are faced with the possibility that there are
many SRE bosonic states with the same thermal Hall conductance
$\kappa_{xy}$, presumably distinguished by a more subtle invariant.
In this paper, we show that this is not the case for $c=16$. The two
SRE bosonic states with $c=16$ edge excitations are equivalent in the bulk:
their partition functions on arbitrary closed manifolds are equal.
However, there are two distinct chiral edge phases of this unique bulk state.
They are connected by an edge reconstruction: a phase transition
must be encountered at the edge in going from one state to the other, but this transition
can occur solely at the edge and the gap need not close in the bulk.
Although we focus on the $c=16$ case, the logic of our analysis readily generalizes.
Therefore, we claim that there is essentially a unique bulk bosonic phase
for each $c=8k$ given by $k$ copies of the so-called $E_8$-state \cite{Kitaev11,Lu12}.
However, there are two distinct {\em fully-chiral} edge phases with $c=16$, twenty-four
with $c=24$, more than ten million with $c=32$, and even more for larger $c$.

One important subtlety arises in our analysis. The two $c=16$
phases do not, initially, appear to be identical. However, when combined
with a trivial insulating phase, the two bulk partition functions can
be mapped directly into each other by a change of variables. This
is a physical realization of the mathematical notion of {\it stable equivalence}.
In general, an effective description of a phase of matter will neglect
many gapped degrees of freedom (e.g., the electrons in inner shells).
However, the sequence of gapped Hamiltonians that interpolates
between two gapped Hamiltonians may involve mixing with these
usually-forgotten gapped degrees of freedom. Therefore, it is natural,
in considering a phase of matter, to allow
an arbitrary enlargement of the Hilbert space by trivial
gapped degrees of freedom (i.e., by SRE phases without gapless
edge excitations). This is useful when, for instance, comparing
a trivial insulating phase with $p$ bands with another trivial insulating
phase with $q>p$ bands. They can be adiabatically connected
if we are allowed to append $q-p$ trivial insulating bands to the latter
system. This notion is also natural when connecting different phases of gapless
edge excitations. The edge of a gapped bulk state will generically have
gapped excitations that we ordinarily ignore. However, they can become 
gapless -- which is a form of edge reconstruction -- and interact
with the other gapless degrees of freedom, driving the edge into a different
phase. However, this does not require any change in the bulk. As we will
see, such a purely edge phase transition connects the two seemingly
different chiral gapped edges with $c=16$. By combining a $c=16$
state with a trivial insulator, we are able to take advantage of
the uniqueness of signature $(8k+n,n)$ even unimodular lattices \cite{Serre73},
from which it follows that the two phases are the same.
This is closely-related to the fact that $T$-duality exchanges
toroidal compactifications of the ${E_8}\times{E_8}$ and
$\text{Spin}(32)/\mathbb{Z}_2$ versions
of the heterotic string, as explained by Ginsparg \cite{Ginsparg87}.

In the remainder of this paper, we describe the equivalence of the
two candidate phases at $k=2$ from two complementary perspectives.
To set the stage, we begin in Section \ref{sec:K-matrix} with a short introduction to the $K$-matrix formalism that we use to describe the phases of matter studied in this paper.
In Section \ref{sec:bulk-equivalence}, we provide a bulk description of the equivalence of the two candidate phases at $k=2$. We then turn to the edge, where we show
that there are two distinct chiral phases of the edge.
We first discuss the fermionic description of the edge modes in Section \ref{sec:fermionic} and then turn to the bosonic description in Section \ref{sec:edge-phase-diagram}.
There is an (purely) edge transition between these two phases. We discuss the
phase diagram of the edge, which is rather intricate, and its relation to the bulk.
In Section \ref{sec:t-vectors}, we summarize how the phase diagram can change when some of the degrees of freedom are electromagnetically charged so that a $U(1)$ symmetry is preserved.
We then conclude in Section \ref{sec:conclusion} and discuss possible generalizations of this picture.

In Appendix \ref{sec:lattices-matrices}, we collect basic definitions and explain the
notation used throughout the text.
In Appendix \ref{sec:dimension-contraction}, we provide some technical details for an argument used in the main text.

\section{K-matrix Formalism}
\label{sec:K-matrix}

\subsection{Chern-Simons Theory}
\label{sec: CStheory}

We will consider $2+1$-dimensional phases of matter
governed by bulk effective field theories of the form:
\begin{eqnarray}
\label{cstheory}
{\cal L} = {1 \over 4 \pi} \epsilon^{\mu \nu \rho} K_{I J} a^I_\mu \partial_\nu a^J_\rho
+ j^{\mu}_I a_\mu^I,
\end{eqnarray}
where $a_\mu^I$, for $I = 1, ..., N$ and $\mu = 0, 1, 2$.
See Refs. \onlinecite{Wen92a} and \onlinecite{wenreview} for a pedagogical introduction to such phases.
$K_{IJ}$ is a symmetric, non-degenerate $N\times N$ integer matrix. 
(Repeated indices should be summed over unless otherwise specified.)
We normalize the gauge fields $a_\mu^I$ and sources $j^{\mu}_I$
so that fluxes that are multiples of $2\pi$ are unobservable
by the Aharonov-Bohm effect. Consequently,
if we take the sources to be given by prescribed non-dynamical
classical trajectories ${x_{m}^\mu}(\tau)$ that serve as sources of $a^I_\mu$ flux,
they must take the form:
\begin{equation}
j^{\mu}_I={\sum_m}{n^{(m)}_I}\delta({x^\mu}-{x_{m}^\mu}(\tau)) {\partial_\tau}x_{m}^\mu,
\end{equation}
for integers ${n^{(m)}_I}$. 
The sum over $m$ is a sum over the possible sources $x_m$.

Therefore, each excitation $m$ of the system is associated with an
integer vector $n_I^{(m)}$. 
These integer vectors can be associated with the points
of a lattice as follows. 
Let $\lambda_a$ for $a=1,\ldots, N$
be the eigenvalues of $(K^{-1})^{IJ}$ with 
$f^I_a$ the corresponding eigenvectors.
We normalize the $f^I_a$ so that 
$(K^{-1})^{IJ} = \eta^{ab} f^I_a f^J_b$
where $\eta^{ab}=\text{sgn}({\lambda_a})\delta^{ab}$.
Now suppose that we view the $f^I_a$ as the components of a
vector ${\bf f}^I \in \mathbb{R}^{N_+ , N_-}$
(i.e., of $\mathbb{R}^{N}$ with a metric
$\eta_{ab}= \text{sgn}({\lambda_a})\,\delta_{ab}$ of signature $(N_+ , N_-)$),
where $K^{-1}$ has $N_+$ positive eigenvalues and $N_-$ negative ones.
In other words, the unit vector ${\bf \hat{x}_a} = (0, ..., 0, 1, 0, ..., 0)^{{\rm tr}}$ with a $1$ in the a-th entry and zeros otherwise is an orthonormal basis of $\mathbb{R}^{N_+ , N_-}$
so that ${\bf \hat{x}_a}\cdot {\bf \hat{x}_b} \equiv ({\bf \hat{x}_a})^c \eta_{c d} ({\bf \hat{x}_b})^d = \eta_{ab}$.
Then we can define ${\bf f}^I \equiv f^I_a {\bf \hat{x}_a}$.
Thus, the eigenvectors ${\bf f}^I$ define a lattice $\Gamma$ in $\mathbb{R}^{N_+ , N_-}$
according to $\Gamma = \{ m_I {\bf f}^I  | {m_I} \in \mathbb{Z}\}$; this lattice determines the allowed excitations of the system \cite{Read90, blockwen90}.

The lattice $\Gamma$ enters directly into the computation of various physical observables.
For example, consider two distinct excitations corresponding to
the lattice vectors ${\bf u} = m_I {\bf f}^I$ and ${\bf v} = n_J {\bf f}^J$ in $\Gamma$.
If one excitation is taken fully around the other, then the resulting wavefunction differs from its original value by the exponential of the Berry's phase $2\pi (K^{-1})^{IJ} {m_I} {n_J} = 2\pi {\bf u}\cdot {\bf v}$.
When the excitations are identical, ${\bf u} = {\bf v}$, a half-braid is sufficient and a phase equal to $\pi {\bf u}\cdot {\bf u}$ is obtained.

Of course, any basis of the lattice $\Gamma$ is equally good; there is nothing special
about the basis ${\bf f}^I$. We can change to a different basis
${\bf f}^I = W^I_{\ J} {\bf \tilde{f}}^J$, where $W \in SL(N,\mathbb{Z})$.
($W$ must have integer entries since it relates one set of lattice vectors
to another. Its inverse must also be an integer matrix since either
set must be able to serve as a basis. But since $\text{det}(W) = 1/\text{det}(W^{-1})$,
$W$ and $W^{-1}$ can both be integer matrices only if $\text{det}(W)=\pm 1$.)
This lattice change of basis can be interpreted as the field redefinitions, ${\tilde a}^I_\mu=W^I_{\ J} a^J_\mu$ and ${\tilde j}_I^{\mu} W^I_{\ J}={j}^{\mu}_J$, in terms of which the Lagrangian (\ref{cstheory}) becomes
\begin{eqnarray}
\label{cstheory-similarity}
{\cal L} = {1 \over 4 \pi} \epsilon^{\mu \nu \rho} {\tilde K}_{I J}
{\tilde a}^I_\mu \partial_\nu {\tilde a}^J_\rho
+ {\tilde j}^{\mu}_I {\tilde a}_\mu^I,
\end{eqnarray}
where $K=W^T {\tilde K} W$. Therefore, two theories are physically identical
if their $K$-matrices are related by such a similarity transformation.

We note that the low energy phases described here may
be further sub-divided according to their coupling to the electromagnetic
field, which is determined by the $N$-component vector $t_I$:
\begin{eqnarray}
\label{cstheory-with-t}
{\cal L} = {1 \over 4 \pi} \epsilon^{\mu \nu \rho} K_{I J} a^I_\mu \partial_\nu a^J_\rho
+ j^{\mu}_I a_\mu^I
- {1 \over 2\pi} \epsilon^{\mu \nu \rho} t_I A_\mu \partial_\nu a_\rho^I.
\end{eqnarray}
It is possible for two theories with the same $K$-matrix to correspond
to different phases if they have different
$t_I$ vectors since they may have different Hall conductances
$\sigma_{xy}=\frac{e^2}{h} (K^{-1})^{IJ} t_I t_J$.
(It is also possible for discrete global symmetries, such as time-reversal,
to act differently on theories with the same $K$-matrix in which case
they can lead to different SPT phases if that symmetry is present.)

In this paper, we will be interested in states of matter in which all
excitations have bosonic braiding properties, i.e., in which any exchange
of identical particles or full braid of distinguishable particles
leads to a phase that is a multiple of $2\pi$.
Hence, we are interested in lattices for which
${\bf f}^I \cdot {\bf f}^J$ is an integer for all $I, J$ and is an even integer
if $I=J$. 
Hence, $K^{-1}$ is a symmetric integer matrix with even entries on the diagonal.
By definition $K$ must also an integer matrix. Since both
$K$ and $K^{-1}$ are integer matrices, their determinant must be $\pm 1$.
Because ${\bf f}^I \cdot {\bf f}^I \in 2\mathbb{Z}$ (no summation on $I$) and
$\text{det}({\bf f}^I \cdot {\bf f}^J)=\pm 1$,
the lattice $\Gamma$ is said to be an even unimodular lattice.

It is convenient to introduce the (dual) vectors $e^a_I = K_{I J} \eta^{a b} f_b^J$.
If, as above, we view the $e^a_I$ as the components of a vector
${\bf e}_I \in \mathbb{R}^{N_+ , N_-}$
according to ${\bf e}_I \equiv e_I^a {\bf \hat{x}_a}$, then
$K_{IJ} = {\bf e}_I \cdot {\bf e}_J$. Moreover,
${\bf e}_I$ is the basis of the dual lattice $\Gamma^*$ defined by
${\bf f}^I \cdot {\bf e}_J = \delta^I_{\ J}$.
Since the lattice $\Gamma$ is unimodular,
it is equal to $\Gamma^*$, up to an $SO(N_+ , N_-)$ rotation, from which
we see that $K$ must be equivalent to $K^{-1}$, up to an
$SL(N,\mathbb{Z})$ change of basis.
(In fact, the required change of basis is provided by the defining relation
$e^a_I = K_{I J} \eta^{a b} f_b^J$.)

Now consider the Lagrangian (\ref{cstheory-rescaled}) on the spatial torus.
For convenience, we assume there are no sources so ${\bf j}^\mu = 0$.
We can rewrite the Lagrangian as
\begin{eqnarray}
\label{cstheory-rescaled}
{\cal L} &=& {1 \over 4 \pi} \epsilon^{\mu \nu \rho} {\bf e}_I \cdot {\bf e}_J
a^I_\mu \partial_\nu a^J_\rho
+ j^{\mu}_I {\bf f}^I \cdot {\bf e}_J a_\mu^J\\
&=& {1 \over 4 \pi} \epsilon^{\mu \nu \rho} 
{\bf a}_\mu \cdot \partial_\nu {\bf a}_\rho
+ {\bf j}^{\mu} \cdot  {\bf a}_\mu,
\end{eqnarray}
where we have defined ${\bf a}_\mu \equiv {\bf e}_I a^I_\mu$ and
${\bf j}^{\mu}\equiv {\bf f}^I j^{\mu}_I$. 
Choosing the gauge ${\bf a}_0 = 0$,
$\partial_i {\bf a}_i = 0$,
the Lagrangian takes the form:
\begin{eqnarray}
\label{cstheory-torus}
{\cal L} = - {1 \over 2 \pi} {\bf a}_1 \cdot \partial_t {\bf a}_2.
\end{eqnarray}
Therefore, ${\bf a}_1$ and ${\bf a}_2$ are canonically conjugate.
Although we have gauge-fixed the theory for small gauge transformations, under a large gauge transformation,
$a^I_k \rightarrow a^I_k + n^I_{(k)}$
where $n^I_{(k)}$ are integers (so that physical observables such as the Wilson loop $e^{i\oint_{C_k} a^I_k}$ about the 1-cycle $C_k$ remains invariant). 
Therefore, we must identify ${\bf a}_j$ and  ${\bf a}_j + n^I_{(k)} {\bf e}_I$
since they are related by a gauge transformation. 

Suppose that we write a ground state wavefunction in the form $\Psi[{\bf a}_1]$.
Then ${\bf a}_1$ will act by multiplication and its canonical conjugate
${\bf a}_2$ will act by differentiation. 
To display the full gauge invariance of the wavefunction, $\Psi[{\bf a}_1]=\Psi[{\bf a}_1 + n^I {\bf e}_I]$, it is instructive
to expand it in the form:
\begin{equation}
\label{eqn:gs-wavefunction}
\Psi[{\bf a}_1] = {\cal N} \sum_{m_I} \Psi_{m_I} e^{2\pi i {m_I} {\bf f}^I \cdot {\bf a}_1}
\end{equation}
where ${m_I}\in \mathbb{Z}$. This is an expansion in eigenstates of
${\bf a}_2$, with the $m_I$ term having the eigenvalue $2 \pi i {m_I} {\bf f}^I$.
However, by gauge invariance, ${\bf a}_1$ takes values in ${\mathbb{R}^N}/\Gamma^*$.
Therefore, we should restrict $m_I$ such that ${m_I} {\bf f}^I$ lies inside
the unit cell of $\Gamma^*$. In other words, the number of ground states
on the torus is equal to the number of sites of $\Gamma$ that lie
inside the unit cell of $\Gamma^*$. This is simply the ratio of the volumes of
the unit cells, $|\text{det}(K)|^{1/2}/|\text{det}(K)|^{-1/2}=|\text{det}(K)|$.
It may be shown that this result generalizes to a ground state
degeneracy $|\text{det}K|^g$ on a genus $g$ surface \cite{Wen90a}. Therefore,
the theories on which we focus in this paper have non-degenerate
ground states on an arbitrary surface, which is another manifestation
of the trivial braiding properties of its excitations.

One further manifestation of the trivial braiding properties of such
a phase's excitations is the bipartite entanglement entropy
of the ground state \cite{kitaevpreskill, levinwenentanglement}. 
If a system with action (\ref{cstheory})
with $j^{\mu}_I = 0$ is divided into two subsystems
$A$ and $B$ and the reduced density matrix $\rho_A$ for subsystem $A$
is formed by tracing out the degrees of freedom of subsystem
$B$, then the von Neumann entropy
$S_A = - {\rm tr}\Big(\rho_A \log(\rho_A)\Big)$ takes the form:
\begin{equation}
S_A = \alpha L - \ln \sqrt{|\text{det}(K)|} + \ldots
\end{equation}
Here, $\alpha$ is a non-universal constant that vanishes
for the action (\ref{cstheory}), but is non-zero if we include
irrelevant sub-leading terms in the action (e.g., Maxwell terms for the gauge fields). 
$L$ is the length of the boundary
between regions $A$ and $B$. 
The $\ldots$ denote terms
with sub-leading $L$ dependence. For the theories that
we will consider in this paper, the second term, which is universal,
vanishes. For this reason, such phases are called ``short-range entangled.''
 
The discussion around Eq. (\ref{eqn:gs-wavefunction}), though
essentially correct as far as the ground state degeneracy is concerned,
swept some subtleties under the rug. A more careful treatment \cite{Belov05}
uses holomorphic coordinates ${\cal\bf a} = {\bf a}_1 + i K\cdot {\bf a}_2$,
in terms of which the wavefunctions are $\vartheta$-functions.
Moreover, the normalization
${\cal N}$ must account for the fact that the wavefunction $\Psi$
is a function only on the space of ${\bf a}_i$ with vanishing field strength (which
the ${\bf a}_0 = 0$ gauge constraint requires),
not on arbitrary ${\bf a}_i$. Consequently, it depends on the modular parameter
of the torus as ${\cal N}= (\eta(\tau))^{-N_+} (\eta(\overline{\tau}))^{-N_-}$
where $N_\pm$ are the number of positive and negative eigenvalues of
$K_{IJ}$; the torus is defined by the parallelogram in the complex plane
with corners at $0$, $1$, $\tau$, $\tau+1$ and opposite sides identified;
and is $\eta(\tau)=q^{\frac{1}{24}}\prod_{n=1}^\infty (1-q^n)$ is the Dedekind
$\eta$ function, where $q=e^{2\pi i \tau}$. Consequently, the ground state wavefunction
transforms non-trivially under the mapping class group of the torus
(i.e., under diffeomorphisms of the torus that are disconnected from
the identity, modulo those that can be deformed to the identity) which
is equal to the modular group $SL(2,\mathbb{Z})$
generated by $S: \tau \rightarrow -1/\tau$ and $T: \tau \rightarrow \tau +1$.
Under $T$, which cuts open the torus along its longitude,
twists one end of the resulting cylinder
by $2\pi$, and then rejoins the two ends of the cylinder to reform the torus,
thereby enacting $\tau \rightarrow \tau +1$, the ground state transforms according
to $\Psi \rightarrow e^{-2\pi i (N_+ - N_-)/24} \, \Psi$.
Therefore, so long as $N_+ - N_- \not \equiv 0\  (\text{mod}\ 24)$, the bulk
is not really trivial.

\subsection{Edge Excitations}

The non-trivial nature of these states is reflected in more
dramatic fashion on surfaces with
a boundary, where there may be gapless edge excitations.
For simplicity, consider the disk $D$ with no sources in
its interior \cite{Elitzur89, Wen92edgereview}.
The action (\ref{cstheory}) is invariant under gauge transformations
$a^I_\mu \rightarrow a^I_\mu + -i({g^I})^{-1}\partial_\mu {g^I}$, where
${g^I}\in [U(1)]^N$, so long as ${g^I}=1$ at the boundary $\partial D$.
In order to fully specify the theory on a disk, we must fix the
boundary conditions.
Under a variation of the gauge fields $\delta{a^J_\mu}$,
the variation of the action $S~=~\int_{\mathbb{R}\times D}L$
(here, $\mathbb{R}$ is the time direction) is
\begin{multline}
\delta S = {1 \over 2 \pi} \int_{\mathbb{R}\times D}
 \delta a^I_\mu \, K_{IJ}
 \epsilon^{\mu \nu \rho}\partial_\nu a^J_\rho \\ + 
{1 \over 4 \pi} \int_{\mathbb{R}\times\partial D}
\epsilon^{\mu \nu r} K_{I J} a^I_\mu \delta a^J_\nu  
\end{multline}
Here $r$ is the radial coordinate on the disk.
The action will be extremized by $K_{IJ}
 \epsilon^{\mu \nu \rho}\partial_\nu a^J_\rho=0$
 (i.e. there won't be extra boundary terms in the equations
 of motion) so long as we take boundary conditions such that
 $\epsilon^{\mu \nu r} K_{IJ} a^I_\mu \delta a^J_\nu = 0$.
We can take boundary condition $K_{IJ} a^I_0 + V_{IJ} a^I_x = 0$,
where $x$ is the azimuthal coordinate.
Here $V_{IJ}$ is a symmetric matrix that is determined by
non-universal properties of the edge such as how sharp it is.
The Lagrangian (\ref{cstheory}) is invariant under
all transformations ${a^J_\mu}(x) \rightarrow {a^J_\mu}(x)
-i({g^J})^{-1}(x) {\partial_\mu} {g^J}(x)$
that are consistent with this boundary condition.
Only those with ${g^J}=1$ at the boundary are gauge symmetries.
The rest are ordinary symmetries of the theory.
Therefore, although all bulk degrees of freedom on the disk are fixed by
gauge invariance and the Chern-Simons constraint, there are local
degrees of freedom at the boundary.

The Chern-Simons constraint $K_{IJ} \epsilon_{ij}\partial_i a^J_j =0$
can be solved by taking $a^I_i = ({U^I})^{-1} \partial_i {U^I}$
or, writing ${U^I}=e^{i\phi}$, $a^I_i = \partial_i \phi$, where $\phi\equiv \phi+2\pi$.
This gauge field is pure gauge everywhere in the interior of the disk
(i.e. we can locally set it to zero in the interior with a gauge transformation),
but it is non-trivial on the boundary because we can only make gauge
transformations that are consistent with the boundary condition.
Substituting this expression
into the action (\ref{cstheory}), we see that the action is a total derivative
which can be integrated to give a purely boundary action:
\begin{equation}
S = \frac{1}{4\pi}\,\int\,dt\,dx\,
\left[{K_{IJ}}\,{\partial_t}{\phi^I}\,{\partial_x}{\phi^J}
- {V_{IJ}}\,{\partial_x}{\phi^I}\,{\partial_x}{\phi^J}\right].
\label{edgetheory}
\end{equation}
The Hamiltonian associated with this action will be positive semi-definite
if and only if ${V_{IJ}}$ has non-negative eigenvalues.
If we define ${\bf X} \equiv {\bf e}_J \phi^J$ or, in components,
$X^a \equiv e^a_J \phi^J$, then we can rewrite this in the form
\begin{equation}
S = \frac{1}{4\pi}\,\int\,dt\,dx\,
\left[\eta_{ab}{\partial_t}{X^a}\,{\partial_x}{X^b}
- {v_{ab}}\,{\partial_x}{X^a}\,{\partial_x}{X^b}\right],
\label{edgetheory2}
\end{equation}
where ${v_{ab}}\equiv {V_{IJ}}{f_a^I} {f_b^J}$. 
We see that the velocity matrix $v_{ab}$ parameterizes density-density interactions between the edge modes.
Note that the fields
$X^a$ satisfy the periodicity conditions
$X^a \equiv X^a + 2\pi e^a_I n^I$ for $n^I \in \mathbb{Z}$.

This theory has $N$ different dimension-$1$ fields $\partial_x \phi^I$.
The theory also has `vertex operators', or exponentials of these
fields that must be consistent with their periodicity conditions:
$e^{i m_I \phi^I}$ or, equivalently, $e^{i m_I {\bf f}^I \cdot {\bf X}}$
or, simply, $e^{i {\bf u}\cdot {\bf X}}=e^{i \eta_{ab} u^a X^b}$
for ${\bf u}\in \Gamma$. They have correlation functions:
\begin{multline}
\label{vertex-op-corr-fcn}
\left\langle e^{i {\bf u}\cdot {\bf X}} e^{-i {\bf u}\cdot {\bf X}}\right\rangle
\\
= \prod_{b=1}^{N_+}\frac{1}{(x-{v_b}t)^{y_b}}
\prod_{b = N_+ +1}^{N}\frac{1}{(x+{v_b}t)^{y_b}}
\end{multline}
In this equation,
$y_b \equiv \sum_{a,c,d,e} u_a S_{ab} \eta_{bc} (S^T)_{cd}\eta_{de} u_e$,
where $S_{ab}$ is an $SO(N)$ matrix that diagonalizes $\eta_{ab}v_{bc}$.
Its first $N_+$ columns are the normalized eigenvectors corresponding to positive
eigenvalues of $\eta_{ab}v_{bc}$ and the next $N_-$ columns are the
normalized eigenvectors corresponding to negative eigenvalues of $\eta_{ab}v_{bc}$.
The velocities ${v_b}$ are the absolute values of the eigenvalues
of $\eta_{ab}v_{bc}$.
Therefore, this operator has scaling dimension
\begin{equation}
\label{vertex-scaling-dim}
\Delta_{\bf u} = \frac{1}{2}\sum_{b=1}^{N} y_b.
\end{equation}
The scaling dimensions of an operator in a non-chiral theory generally depend upon the velocity matrix $v_{ab}$.
For a fully chiral edge, however, $\eta_{ab}=\delta_{ab}$,
so $\Delta_{\bf u} = \frac{1}{2} |{\bf u}|^2$.

If the velocities all have the same absolute value, $|{v_a}|=v$ for all $a$,
then the theory is a conformal field theory with right and left Virasoro
central charges $c=N_+$ and $\overline{c}=N_-$. Consequently,
we can separately rescale the right- and left-moving coordinates:
$(x-vt) \rightarrow \lambda(x-vt)$ and $(x+vt) \rightarrow \lambda' (x+vt)$.
The field $\partial_x X^a$ has right and left scaling
dimension $(1,0)$ for $a=1, 2,\ldots, N_+$
and dimension $(0,1)$ for $a=N_+ +1,\ldots, N$. Meanwhile,
$e^{i {\bf u}\cdot {\bf X}}$ has scaling dimension:
\begin{equation}
\label{vertex-RL-scaling-dim}
(\Delta^R_{\bf u},\Delta^L_{\bf u}) =
(\frac{1}{2}\sum_{b=1}^{N_+} y_b,\frac{1}{2}\sum_{b=N_+ + 1}^N y_b).
\end{equation}
which simplifies, for the case of a fully chiral edge, to
$(\Delta^R_{\bf u},\Delta^L_{\bf u}) = (\frac{1}{2} {\bf u} \cdot {\bf u},0)$.

In a slight abuse of terminology, we will call the state of matter
described by Eq. (\ref{cstheory}) in the bulk and Eq. (\ref{edgetheory})
on the edge a $c=N_+$, $\overline{c}=N_-$ bosonic SRE phase.
In the case of fully chiral theories that have $\overline{c}=0$,
we will sometimes simply call them $c=N$ bosonic SRE phases.
Strictly speaking, the gapless edge excitations are only
described by a conformal field theory when the velocities are all
equal.
However, we will continue to use this terminology even when the
velocities are not equal, and we will use it to refer to both the bulk
and edge theories.

In the case of a $c>0$, $\overline{c}=0$ bosonic SRE
phase, all possible perturbations of the edge effective field theory
Eq. (\ref{edgetheory}) -- or, equivalently, Eq. (\ref{edgetheory2}) -- are chiral.
Since such perturbations cannot open a gap, completely chiral edges
are stable. A non-chiral edge may have a vertex operator $e^{i {\bf u}\cdot {\bf X}}$ with
equal right- and left-scaling dimensions. If its total scaling dimension
is less than $2$, it will be relevant and can open a gap at weak coupling. More generally,
we expect that a bosonic SRE will have stable gapless edge excitations
if $c-\overline{c}>0$. Some of the degrees of freedom of the theory
(\ref{edgetheory}) will be gapped out, but some will remain gapless
in the infrared (IR) limit and the remaining degrees of freedom will be fully chiral with
$c_{IR}=c-\overline{c}$ and $\overline{c}_{IR}=0$. Therefore, even if
such a phase is not, initially, fully-chiral, the degrees of freedom that
remain stable to arbitrary perturbations is fully chiral.
Therefore, positive-definite even unimodular lattices correspond to
$c>0$, $\overline{c}=0$ bosonic SRE phases with stable chiral edge excitations,
in spite of the absence of anyons in the bulk.

\subsection{The Cases $c-\overline{c}=0,8,16$}

Positive-definite even unimodular lattices only exist in dimension
$8k$ for integer $k$, \cite{Serre73} so bosonic SRE phases with stable chiral edge excitations
must have $c=8k$. There is a unique positive-definite
even unimodular lattice in dimension $8$, up to
an overall rotation of the lattice. There are two positive-definite even unimodular lattices in dimension $16$; there are $24$ in dimension $24$; there are more than
$10^7$ in dimension $32$; and even more in higher dimensions.
If we relax the condition of positive definiteness, then there are
even unimodular lattices in all even dimensions; there is a unique
one with signature $(8k+n,n)$ for $n\geq 1$.

In dimension-2, the unique even unimodular lattice in $\mathbb{R}^{1,1}$, which we will
call $U$, has basis vectors ${\bf e}_1={1 \over r} ({\bf \hat{x}_1} + {\bf \hat{x}_2})$, ${\bf e}_2 = {r \over 2} ({\bf \hat{x}_1} - {\bf \hat{x}_2})$,
and the corresponding $K$-matrix is:
\begin{equation}
\label{eqn:K_U}
{K_U} = {\bf e}_1 \cdot {\bf e}_2 = \begin{pmatrix}
0 & 1\\
1 & 0
\end{pmatrix}.
\end{equation}
This matrix has signature $(1,1)$.
(Within this discussion, $r$ is an arbitrary parameter.
It will later develop a physical meaning and play an important role in the phase transition we describe.)
The even unimodular lattice of signature $(n,n)$ has a
block diagonal $K$-matrix with $n$ copies of $K_U$ along
the diagonal:
\begin{equation}
K_{U\oplus U \oplus \ldots \oplus U} = \begin{pmatrix}
{K_U} & 0 & 0 & \ldots\\
0 & {K_U} & 0 &\\
0 & 0 & {K_U}& \\
\vdots & & & \ddots
\end{pmatrix}.
\end{equation}

The unique positive definite even unimodular lattice in dimension-$8$
is the lattice generated by the roots of the Lie algebra of $E_8$.
We call this lattice $\Gamma_{E_8}$.
The basis vectors for $\Gamma_{E_8}$ are given in Appendix \ref{sec:lattices-matrices},
and the corresponding $K$-matrix takes the form:
\begin{eqnarray}
K_{E_8} = \begin{pmatrix}
2 & -1 & 0 & 0 & 0 & 0 & 0 & 0 \cr
-1 & 2 & -1 & 0 & 0 & 0 & -1 & 0 \cr
0 & -1 & 2 & -1 & 0 & 0 & 0 & 0 \cr
0 & 0 & -1 & 2 & -1 & 0 & 0 & 0 \cr
0 & 0 & 0 & -1 & 2 & -1 & 0 & 0 \cr
0 & 0 & 0 & 0 & -1 & 2 & 0 & 0 \cr
0 & -1 & 0 & 0 & 0 & 0 & 2 & -1 \cr
0 & 0 & 0 & 0 & 0 & 0 & -1 & 2 \cr
\end{pmatrix}.
\end{eqnarray}

The two positive-definite even unimodular lattices in dimension $16$
are the lattices generated by the roots of $E_8\times E_8$
and $\text{Spin}(32)/\mathbb{Z}_2$. (The latter means that
a basis for the lattice is given by the roots of $SO(32)$, but
with the root corresponding to the vector
representation replaced by the weight of one of the
spinor representations.)
We will call these lattices $\Gamma_{E_8}\oplus\Gamma_{E_8}$
and $\Gamma_{\text{Spin}(32)/\mathbb{Z}_2}$.
They are discussed further in Appendix \ref{sec:lattices-matrices}.
The corresponding $K$-matrices take the form:
\begin{equation}
K_{E_8\times E_8} = \begin{pmatrix}
K_{E_8} & 0 \\
0 & K_{E_8}\\
\end{pmatrix},
\end{equation}
(for later convenience, we permute the rows and columns of the second copy
of $E_8$ in Eq. (\ref{eqn:K-E8-E8}) so that it looks superficially different from
the first ) and
\begin{widetext}
\begin{equation}
K_{\text{Spin}(32)/\mathbb{Z}_2} =
\left(
\begin{array}{cccccccccccccccc}
2 & -1 & 0 & 0 & 0 & 0 & 0 & 0 & 0 & 0 & 0 & 0 & 0 & 0 & 0 & 0 \\
 -1 & 2 & -1 & 0 & 0 & 0 & 0 & 0 & 0 & 0 & 0 & 0 & 0 & 0 & 0 & 0 \\
 0 & -1 & 2 & -1 & 0 & 0 & 0 & 0 & 0 & 0 & 0 & 0 & 0 & 0 & 0 & 0 \\
 0 & 0 & -1 & 2 & -1 & 0 & 0 & 0 & 0 & 0 & 0 & 0 & 0 & 0 & 0 & 0 \\
 0 & 0 & 0 & -1 & 2 & -1 & 0 & 0 & 0 & 0 & 0 & 0 & 0 & 0 & 0 & 0 \\
 0 & 0 & 0 & 0 & -1 & 2 & -1 & 0 & 0 & 0 & 0 & 0 & 0 & 0 & 0 & 0 \\
 0 & 0 & 0 & 0 & 0 & -1 & 2 & -1 & 0 & 0 & 0 & 0 & 0 & 0 & 0 & 0 \\
 0 & 0 & 0 & 0 & 0 & 0 & -1 & 2 & -1 & 0 & 0 & 0 & 0 & 0 & 0 & 0 \\
 0 & 0 & 0 & 0 & 0 & 0 & 0 & -1 & 2 & -1 & 0 & 0 & 0 & 0 & 0 & 0 \\
 0 & 0 & 0 & 0 & 0 & 0 & 0 & 0 & -1 & 2 & -1 & 0 & 0 & 0 & 0 & 0 \\
 0 & 0 & 0 & 0 & 0 & 0 & 0 & 0 & 0 & -1 & 2 & -1 & 0 & 0 & 0 & 0 \\
 0 & 0 & 0 & 0 & 0 & 0 & 0 & 0 & 0 & 0 & -1 & 2 & -1 & 0 & 0 & 0 \\
 0 & 0 & 0 & 0 & 0 & 0 & 0 & 0 & 0 & 0 & 0 & -1 & 2 & -1 & -1 & 0 \\
 0 & 0 & 0 & 0 & 0 & 0 & 0 & 0 & 0 & 0 & 0 & 0 & -1 & 2 & 0 & 0 \\
 0 & 0 & 0 & 0 & 0 & 0 & 0 & 0 & 0 & 0 & 0 & 0 & -1 & 0 & 2 & -1 \\
 0 & 0 & 0 & 0 & 0 & 0 & 0 & 0 & 0 & 0 & 0 & 0 & 0 & 0 & -1 & 4 \\
\end{array}
\right).
\end{equation}
\end{widetext}

The even unimodular lattice with signature $(8+n,n)$
has K-matrix:
\begin{equation}
K_{{E_8}\oplus U\oplus \ldots\oplus U} = \begin{pmatrix}
K_{E_8} & 0 & 0 &\ldots\\
0 & U & 0 & \\
0 & 0 & U &\\
\vdots & & & \ddots\\
\end{pmatrix}.
\end{equation}
The even unimodular lattice with signature $(16+n,n)$
has K-matrix:
\begin{equation}
\label{e8-e8-u}
K_{{E_8}\times{E_8}\oplus U \oplus \ldots\oplus U} = \begin{pmatrix}
K_{E_8} & 0 & 0 &\ldots\\
0 & K_{E_8} & 0 & \\
0 & 0 & U &\\
\vdots & & & \ddots\\
\end{pmatrix}.
\end{equation}
These lattices are unique, so the matrix,
\begin{equation}
\label{so(32)-u}
K_{{\text{Spin}(32)/\mathbb{Z}_2}\oplus U\oplus \ldots\oplus U} = \begin{pmatrix}
K_{\text{Spin}(32)/\mathbb{Z}_2} & 0 & \ldots\\
0 & U &  \\
\vdots & & \ddots\\
\end{pmatrix},
\end{equation}
is equivalent to (\ref{e8-e8-u}) under an $SL(16+2n,\mathbb{Z})$
basis change. This fact will play an important role in the sections
that follow.

\section{Equivalence of the Two $c=16$ Bosonic SRE Phases}
\label{sec:bulk-equivalence}

In the previous section, we saw that two theories of the form
(\ref{cstheory}) with different $N\times N$ $K$-matrices are equivalent if
the two $K$-matrices are related by an $SL(N,\mathbb{Z})$ transformation
or, equivalently, if they correspond to the same lattice.
But if two $K$-matrices are not related by an $SL(N,\mathbb{Z})$ transformation, is there a more general notion that may relate the theories? 
A more general notion might be expected if the difference in the number of positive and negative eigenvalues of the two $K$-matrices coincide.
Consider, for instance, the case
of an ${N_1}\times {N_1}$ $K$-matrix and an ${N_2}\times {N_2}$ $K$-matrix
with ${N_1}<{N_2}$. 
Could there be a relation between them, even though they clearly
cannot be related be related by an $SL({N_1},\mathbb{Z})$ or
$SL({N_2},\mathbb{Z})$ similarity transformation?

The answer is yes, for the following reason. Consider the theory
associated with $K_U$, defined in Eq. (\ref{eqn:K_U}). Its partition
function is equal to $1$ on an arbitrary $3$-manifold, $M_3$, as was shown
in Ref. \onlinecite{Witten03}:
\begin{equation}
Z({M_3}) \equiv \int {\cal D}{a_I}
e^{i\int {1 \over 4 \pi} \epsilon^{\mu \nu \rho}
({K_U})_{I J} a^I_\mu \partial_\nu a^J_\rho}
= 1.
\end{equation}
One manifestation of the triviality of this theory in the bulk
is that it transforms trivially under modular transformations, as we saw earlier.
Furthermore, a state with this $K$-matrix can be smoothly connected
to a trivial insulator by local unitary transformations if no symmetries are maintained \cite{Chen11a}.
We shall not do so here, but it is important to note
that, if we impose a symmetry on the theory, then we can guarantee
the existence of gapless (non-chiral) excitations that live at the edge of the
system \cite{Chen11a,Lu12}.
(We emphasize that we focus, in this section, on the bulk and, in this
paper, on properties that do not require symmetry.)

Therefore, we can simply replace it with a theory with no degrees
of freedom. We will denote such a theory by $K=\emptyset$
to emphasize that it is a $0\times 0$ $K$-matrix in a theory
with $0$ fields and {\em not} a theory with a $1\times 1$ $K$-matrix
that vanishes. Similarly, the partition function for a theory with
arbitrary $K$-matrix $K_A$ on any $3$-manifold $M_3$
is equal to the partition function of $K_{A\oplus U}$
\begin{multline}
\int {\cal D}{a_I}
e^{\frac{i}{4\pi} \int \epsilon^{\mu \nu \rho}
({K_A})_{I J} a^I_\mu \partial_\nu a^J_\rho} = \\
\int {\cal D}{a_I}\,{\cal D}{a'_I} \Bigl[
e^{\frac{i}{4\pi} \int \epsilon^{\mu \nu \rho}
({K_A})_{I J} a^I_\mu \partial_\nu a^J_\rho} \times\\
e^{\frac{i}{4\pi} \int \epsilon^{\mu \nu \rho}
({K_U})_{I J} {a'}^I_\mu \partial_\nu {a'}^J_\rho}\Bigr]\\
= \int {\cal D}{a_I}
e^{\frac{i}{4\pi} \int \epsilon^{\mu \nu \rho}
(K_{A\oplus U})_{I J} a^I_\mu \partial_\nu a^J_\rho}
\end{multline}
Therefore, all of the theories corresponding to even, unimodular lattices
of signature $(n,n)$ are, in fact, equivalent when there is no symmetry preserved. There is just a single
completely trivial gapped phase. We may choose to describe it
by a very large $K$-matrix (which is seemingly perverse), but it is still the same phase.
Moreover, any phase associated with a $K$-matrix can equally
well be described by a larger $K$-matrix to which we have added
copies of $K_U$ along the block diagonal. This is an expression
of the physical idea that no phase transition will be encountered in going
from a given state to one in which additional trivial, gapped degrees of
freedom have been added. Of course, in this particular case, we have
added zero local degrees of freedom to the bulk and we have not enlarged
the Hilbert space at all. So it is an even more innocuous operation.
However, when we turn to the structure of edge excitations,
there will be more heft to this idea.

At a more mathematical level, the equivalence of
these theories is related to the notion of ``stable equivalence'', according to
which two objects are the same if they become isomorphic
after augmentation by a ``trivial'' object.
In physics, stable equivalence has been used in the K-theoretic classification
of (non-interacting) topological insulators \cite{Kitaev09}.
In the present context, we will be comparing gapped phases
and the trivial object that may be added to either phase is a
topologically-trivial band insulator.  
Heuristically, stable equivalence says that we may add some number of topologically-trivial bands to our system in order to effectively enlarge the parameter space and, thereby, allow a continuous interpolation between two otherwise different states.

We now turn to the two $c=16$ bosonic SRE phases.
Their bulk effective field theories are of the form of
Eq. (\ref{cstheory}) with $K$-matrices given by
$K_{E_8 \times E_8}$ and $K_{{\rm Spin}(32)/{\mathbb Z}_2}$.
Their bulk properties are seemingly trivial. But not
entirely so since, as we noted in Section \ref{sec:K-matrix}, they transform
non-trivially under modular transformations.

These two non-trivial theories are, at first glance, distinct.
They are associated with different lattices. For instance,
$\Gamma_{E_8}\oplus\Gamma_{E_8}$ is the direct sum of
two $8$-dimensional lattices while
$\Gamma_{\text{Spin}(32)/\mathbb{Z}_2}$ is not.
The two $K$-matrices are
not related by an $SL(16,\mathbb{Z})$ transformation.

Suppose, however, that we consider the $K$-matrices
$K_{E_8 \times E_8} \oplus U$ and $K_{{\rm Spin}(32)/\mathbb{Z}_2} \oplus U$
which describe "enlarged" systems. (We use quotation marks because,
although we now have theories with $18$ rather than $16$ gauge fields,
the physical Hilbert space has not been enlarged.)
These $K$-matrices are, in fact, related by an $SL(18,\mathbb{Z})$
transformation:
\begin{equation}
\label{ginspargcon}
W_G^{T} \, K_{{\rm Spin}(32)/{\mathbb Z}_2 \oplus U} \, W_G =  K_{E_8 \times E_8 \oplus U},
\end{equation}
where $W_G$ is given by:
\begin{widetext}
\begin{eqnarray}
W_G = \left(
\begin{array}{cccccccccccccccccc}
-2 & 1 & 0 & 0 & 0 & 0 & 0 & 0 & 0 & 0 & 0 & 0 & 0 & 0 & 0 & 0 & 0 & 0  \\
-3 & 0 & 1 & 0 & 0 & 0 & 1 & 0 & 0 & 0 & 0 & 0 & 0 & 0 & 0 & 0 & 0 & 0 \\
-4 & 0 &0 & 1 & 0 & 0 & 2 & 0 & 0 & 0 & 0 & 0 & 0 & 0 & 0 & 0 & 0 & 0 \\
-5 & 0 & 0& 0 & 1 & 0 & 3 & 0 & 0 & 0 & 0 & 0 & 0 & 0 & 0 & 0 & 0 & 0 \\
-6 & 0 & 0 & 0 & 0 & 1 & 4 & 0 & 0 & 0 & 0 & 0 & 0 & 0 & 0 & 0 & 0 & 0 \\
-7 & 0 & 0 & 0 & 0 & 0 & 5& 0 & 0 & 0 & 0 & 0 & 0 & 0 & 0 & 0 & 0 & 0 \\
-8 & 0 & 0 & 0 & 0 & 0 & 6 & 0 & 0 & 0 & 0 & 0 & 0 & 0 & 0 & 0 & 0 & -1 \\
-9 & 0 & 0 & 0 & 0 & 0 & 7& 0 & 0 & 0 & 0 & 0 & 0 & 0 & 0 & 0 & 1 & -1 \\
-10 & 0 & 0 & 0 & 0 & 0 & 8 & 0 & 1 & 0 & 0 & 0 & 0 & 0 & 0 & 0 & 2 & -2 \\
-11 & 0 & 0 & 0 & 0 & 0 & 9 & 0 & 0 & 1 & 0 & 0 & 0 & 0 & 0 & 0 & 3 & -3 \\
-12 & 0 & 0 & 0 & 0 & 0 & 10 & 0 & 0 & 0 & 1 & 0 & 0 & 0 & 0 & 0 & 4 & -4 \\
-13 & 0 & 0 & 0 & 0 & 0 & 11 & 0 & 0 & 0 & 0 & 1 & 0 & 0 & 0 & 0 & 5 & -5 \\
-14 & 0 & 0 & 0 & 0 & 0 & 12 & 0 & 0 & 0 & 0 & 0 & 1 & 0 & 0 & 0 & 6 & -6 \\
-7 & 0 & 0 & 0 & 0 & 0 & 6 & 0 & 0 & 0 & 0 & 0 & 0 & 1 & 0 & 0 & 3 & -3 \\
-8 & 0 & 0 & 0 & 0 & 0 & 7 & 0 & 0 & 0 & 0 & 0 & 0 & 0 & 1 & 0 & 4 & -4 \\
-2 & 0 & 0 & 0 & 0 & 0 & 2 & 0 & 0 & 0 & 0 & 0 & 0 & 0 & 0& 1 & 2 & -2 \\
0 & 0 & 0 & 0 & 0 & 0 & 1 & -1 & 0 & 0 & 0 & 0 & 0 & 0 & 0 & 1 & 2 & -2 \\
0 & 0 & 0& 0 & 0 & 0 & 0 & -1 & 0 & 0 & 0 & 0 & 0 & 0 & 0 & -1 & -2 & 2 \\
\end{array}
\right).
\end{eqnarray}
\end{widetext}
We will explain how $W_G$ is derived in Section \ref{sec:edge-phase-diagram}.
Here, we focus on its implication:
these two theories are equivalent on an arbitrary
closed manifold. There is a {\rm unique} bulk $c=16$
bosonic SRE phase of matter.
However, there appear to be two possible distinct effective field theories
for the edge of this unique bulk phase, namely the theories
(\ref{edgetheory}) with $K_{E_8 \times E_8}$ and
$K_{{\rm Spin}(32)/{\mathbb Z}_2}$. In the next section, we explain the relation
between these edge theories.

\section{Fermionic Representations of the Two $c=16$ SRE Bosonic Phases}
\label{sec:fermionic}

In Section \ref{sec:bulk-equivalence}, we saw that there is a unique bulk $c=16$
bosonic SRE phase of matter. We now turn our attention to the
two corresponding edge effective field theories, namely
Eq. (\ref{edgetheory}) with $K_{IJ}$ given by either $K_{E_8 \times E_8}$ or
$K_{{\rm Spin}(32)/{\mathbb Z}_2}$. These two edge theories
are distinct, although the difference is subtle.
To understand this difference, it is useful to consider fermionic
representations \cite{Gross85,Green87} of these edge theories.

Consider 32 free chiral Majorana fermions:
\begin{equation}
\label{eqn:fermionic}
S = \int dx d\tau \, {\psi_j}\left(-\partial_\tau + {v_a}i\partial_x\right){\psi_j},
\end{equation}
where $j=1,\ldots,32$. If the velocities $v_a$ are all the same, then
this theory naively has $SO(32)$ symmetry, up to a choice of boundary conditions. 
We could imagine such
a $1+1$-dimensional theory as the edge of a $32$-layer system of electrons,
with each layer in a spin-polarized $p+ip$ superconducting state.
We will assume that the order parameters in the different layers
are coupled by inter-layer Josephson tunneling so that the superconducting
order parameters are locked together. Consequently, 
if a flux $hc/2e$ vortex passes through one of the layers,
it must pass through all 32 layers.
Then all $32$ Majorana fermion edge modes have the same boundary conditions.
When two vortices in a single-layer spin-polarized $p+ip$ superconducting state
are exchanged, the resulting phase is $e^{-i\pi/8}$ or $e^{3i\pi/8}$,
depending on the fusion channel of the vortices (i.e., the fermion parity
of the combined state of their zero modes). Therefore, a vortex
passing through all $32$ layers (which may be viewed as a composite
of $32$ vortices, one in each layer) is a boson. These bosons
carry $32$ zero modes, so there are actually $2^{16}$ states of such
vortices -- $2^{15}$ if we require such a vortex to have even fermion parity.
(Of course, the above construction only required 16 layers if our goal was to construct the minimal dimension SRE chiral phase of bosons. \cite{Kitaev11})

Now suppose that such vortices condense. (Without loss of generality,
we suppose that the vortices are in some particular internal state
with even fermion parity.) Superconductivity is destroyed and the system
enters an insulating phase. Although individual fermions are confined since
they acquire a minus sign in going around a vortex, a pair of fermions,
one in layer $i$ and one in layer $j$, is an allowed excitation.
The dimension-$1$ operators in the edge theory are of the form
$i{\psi_i} {\psi_j}$ where
$1\leq i < j \leq 32$.
There are $\frac{1}{2}\cdot 32\cdot 31 = 496$ such operators.
We may choose $i\psi_{2a-1}\psi_{2a}$, with $a=1, 2, \ldots,16$ as a maximal commuting
subset, i.e. as the Cartan subalgebra of $SO(32)$.
The remaining $480$ correspond to the vectors of $(\text{length})^2 = 2$
in the lattice $\Gamma_{16}$. To see this, it is useful to bosonize the
theory (\ref{eqn:fermionic}). 
We define the Dirac fermions $\Psi_I \equiv \psi_{2a-1} + i \psi_{2a}$,
with $a=1, 2, \ldots,16$
and represent them with bosons: $\Psi_I = e^{i X_a}$. Then the Cartan subalgebra
consists of the $16$ dimension-$1$ operators $\partial{X_a}$.
The operators $e^{i {\bf v}\cdot {\bf X}}$ with ${\bf v} \in \Gamma_{SO(32)} \subset \Gamma_{{\rm Spin}(32)/\mathbb{Z}_2}$
and $|{\bf v}|^2 = 2$ correspond to the vectors of $(\text{length})^2 = 2$ in the
$SO(32)$ root lattice: $\pm{\bf \hat{x}_a} \pm {\bf \hat{x}_b}$ with
$1\leq a < b \leq 16$. In the fermionic language, we see that the relevant perturbations
of $i\psi_i \psi_k$ can be gauged away with a spatially-dependent
$SO(32)$ rotation and, therefore, do not affect the basic physics
of the state.

To complete the description of the ${\rm Spin}(32)/\mathbb{Z}_2$ theory, recall that a vortex in a single layer braids non-trivially with the composite vortex that condenses.
Such single vortices are confined after condensation of the composite. 
Therefore, it is impossible to change the boundary conditions of just one of the fermions $\psi_i$ by inserting a single vortex into the bulk; all of the fermions must have the same boundary conditions.
The fermion boundary conditions can be changed from anti-periodic to periodic
by the operator $e^{i {\bf \mu}_s \cdot {\bf X}}=\exp(i({X_1} + {X_2} + \ldots + X_{16})/2)$,
where ${\bf \mu}_s$ is the weight of one of the spinor representations of
$SO(32)$. This is a dimension-$2$ operator.

Note that the group $\text{Spin}(32)$ is a double-cover of $SO(32)$ that has spinor representations. 
By disallowing one of the spinor representations and the vector representation (i.e., the odd fermion parity sector), the theory is associated with
$\text{Spin}(32)/\mathbb{Z}_2$ but the $\mathbb{Z}_2$ that is modded
out is not the the $\mathbb{Z}_2$ that leads back to $SO(32)$.
Thus, it is the inclusion of ${\bf \mu}_s$ along with the vectors ${\bf v}$ of $SO(32)$ mentioned above that is essential to the description of the fermionic representation of the ${\rm Spin}(32)/\mathbb{Z}_2$ theory. 
If we had chosen not to include ${\bf \mu}_s$, i.e., if we had not condensed the composite vortex, the resulting theory would have had topological order with a torus ground state degeneracy equal to four.
(The $SO(32)$ root lattice has unit cell volume equal to four while the unit cell volume of the ${\rm Spin}(32)/\mathbb{Z}_2$ lattice is unity.)

Now suppose that the first $16$ layers are coupled by interlayer
Josephson tunneling so that their order parameters are locked
and the remaining $16$ layers are coupled similarly,
but the first $16$ layers are not coupled to the remaining $16$.
Then there are independent vortices in the first $16$ layers
and in the remaining $16$ layers. Suppose that both types
of vortices condense. Each of these $16$-vortex composites is a boson, and
superconductivity is again destroyed. Individual fermions are again confined
and, moreover, the fermion parity in each half of the system must be even.
Therefore, the allowed dimension-$1$ operators in the theory are
$i{\psi_i} {\psi_j}$ with $1\leq i < j \leq 16$ or $17\leq i < j \leq 32$.
There are $2\cdot \frac{1}{2}\cdot 16\cdot 15 = 240$ such dimension-$1$
operators. As above, $16$ of them correspond to the Cartan subalgebra.
The other $224$ correspond to lattice vectors
$e^{i {\bf v}\cdot {\bf X}}$ with
${\bf v} = \pm {\bf \hat{x}_a} \pm {\bf \hat{x}_b}$ and
$1\leq a < b \leq 8$ or $9\leq a < b \leq 16$.
Unlike in the case of $\text{Spin}(32)/\mathbb{Z}_2$, the boundary-condition
changing operators $\exp(i(\pm{X_1} \pm {X_2} \ldots \pm X_{8})/2)$
and $\exp(i(\pm X_{9} \pm X_{10} \ldots \pm X_{16})/2)$
are dimension-$1$ operators. There are $2\cdot 2^7 = 256$ such operators
with even fermion parity in each half of the system (i.e., an even number of
$+$ signs in the exponential). The corresponding vectors
${\bf v} = (\pm {\bf \hat{x}_1} \pm {\bf \hat{x}_2} \ldots \pm {\bf \hat{x}_8})/2$
and ${\bf v} = (\pm {\bf \hat{x}_{9}} \pm {\bf \hat{x}_{10}} \ldots \pm {\bf \hat{x}_{16}})/2$
with an even number of $+$ signs together with ${\bf v} = \pm {\bf \hat{x}_a} \pm {\bf \hat{x}_b}$
are the $480$ different $(\text{length})^2 = 2$ vectors in the
${E_8}\times {E_8}$ root lattice. Consequently, this is the
fermionic representation of the ${E_8}\times {E_8}$ theory.

It is unclear, from this fermionic description, how to adiabatically connect
the two bulk theories. The most obvious route between them, starting
from the ${E_8}\times {E_8}$ theory, is to restore superconductivity,
couple the order parameters of the two sets of $16$ layers, and then
condense $32$-layer vortices to destroy superconductivity again.
This route takes the system across three phase transitions while the
analysis in the previous section showed that they are, in fact, the same phase
and, therefore, it should be possible to go from one to the other without
crossing any bulk phase boundaries.

As we saw above, there are $480$ vectors ${\bf u}$ with $|{\bf u}|^2 = 2$
in both $\Gamma_{E_8 \times E_8}$ and
$\Gamma_{{\rm Spin}(32)/\mathbb{Z}_2}$.
In fact, a result of Milnor\cite{Milnor1964} (related to hearing the shape of a drum) states
that the two lattices have the same number of vectors of {\it all lengths}:
for every ${\bf u} \in \Gamma_{E_8 \times E_8}$, there is a unique partner
${\bf v} \in \Gamma_{{\rm Spin}(32)/\mathbb{Z}_2}$ such that $|{\bf v}|^2 = |{\bf u}|^2$.
(See Ref. \onlinecite{Green87} for an elegant presentation of this fact following Ref. \onlinecite{Serre73}.)
Therefore, the $E_8 \times E_8$ and ${\rm Spin}(32)/\mathbb{Z}_2$
edge theories have identical spectra of operator scaling dimensions
$\Delta_{\bf u}=\frac{1}{2}|{\bf u}|^2$. Thus, it is impossible
to distinguish these two edge theories by measuring
the possible exponents associated with
two-point functions. However, in the fermionic realization described above,
consider one of the $496$ dimension-$1$ operators, which we will
call $J_i$, $i=1,2,\ldots, 496$. They are given by
$\partial X_a$ and $e^{i {\bf u}\cdot {\bf X}}$
with $|{\bf u}|^2 = 2$ for ${\bf u} \in \Gamma_{E_8 \times E_8}$ or
$\Gamma_{{\rm Spin}(32)/\mathbb{Z}_2}$.
In the limit that all of the velocities are equal, these are conserved currents
corresponding to the $496$ generators of either $E_8 \times E_8$
or ${\rm Spin}(32)/\mathbb{Z}_2$, but we will use the notation $J_i$
even when the velocities are not equal.
It is clear that, in the ${\rm Spin}(32)/\mathbb{Z}_2$ phase, there are
$J_i$s that involve both halves of the system, but not in the
$E_8 \times E_8$ phase. In other words, in the ${\rm Spin}(32)/\mathbb{Z}_2$ phase,
there are two-point functions involving both halves of the system
that decay as $\langle J_{i}({x},0) J_{i}(0,0) \rangle \propto 1/x^2$.
In the $E_8 \times E_8$ phase, such operators $J_{i}$ only exist
acting entirely within the top half or the bottom half of the system.

Moreover, the $n$-point functions for $n \geq 3$ of the two theories can be different.
Consider the following $4$-point function in our $32$-layer model,
\begin{equation}
\label{eqn:four-point}
\langle J_{i_1}({x_1},{t_1}) J_{i_2}({x_2},{t_2})
J_{i_3}({x_3},{t_3}) J_{i_4}({x_4},{t_4}) \rangle_c,
\end{equation}
where the subscript $c$ denotes a connected correlation function,
and $J_{i_1}$ acts within the first $16$ layers and $J_{i_2}$ within the
second $16$ layers. In the $E_8 \times E_8$ theory,
this correlation function vanishes for all
choices of ${i_3}, {i_4}$ because there are no
dimension-$1$ operators that
act on both halves of the system, i.e. within both
the first $16$ layers and the second $16$ layers.
On the other hand, in the ${\rm Spin}(32)/\mathbb{Z}_2$ theory,
there will always be choices of ${i_3}, {i_4}$ such that
the connected correlation function is non-zero:
if $J_{i_1}=i\psi_k \psi_l$ and $J_{i_2}=i\psi_m \psi_n$
with $1\leq k < l \leq 16$ and $17\leq m < n \leq 32$
then the connected correlation function is non-zero
for $J_{i_3}=i\psi_k \psi_m$ and $J_{i_4}=i\psi_l \psi_n$.
Such a correlation function (\ref{eqn:four-point}) corresponds
to a measurement of a current $J_{i_1}$ in the top half of the system
in response to a probe that couples to $J_{i_2}$ in the bottom half
of the system. While such a measurement will give a vanishing result
in the absence of other perturbations, it will give a non-vanishing
result in the ${\rm Spin}(32)/\mathbb{Z}_2$ theory in the presence of
perturbations that couple to $J_{i_3}$ and $J_{i_4}$. In other words,
it is a measurement of $J_{i_1}$ to linear order in external fields that couple to
$J_{i_2}$, $J_{i_3}$, and $J_{i_4}$.

Of course, in some other physical realization
it may be more difficult to divide these currents into
a `top half' and a `bottom half', but there will always be correlation
functions that distinguish the two edge theories.

\section{Phase Diagram of the $c-\overline{c}=16$ Edge.}
\label{sec:edge-phase-diagram}

Since there is a unique bulk $c=16$ bosonic SRE phase of matter,
the two different edge theories corresponding to $K_{E_8 \times E_8}$ or
$K_{{\rm Spin}(32)/\mathbb{Z}_2}$ must be different edge phases that can
occur at the boundary of the same bulk phase. For this scenario to hold,
it must be the case that the transition between these two edge theories
is purely an edge transition -- or, in other words, an ``edge reconstruction'' --
that can occur without affecting the bulk. Such a transition can occur as follows.
The gapless modes in the effective theory (\ref{edgetheory}) are the
lowest energy excitations in the system. However, there will generically
be gapped excitations at the edge of the system that we usually ignore.
So long as they remain gapped, this is safe. However, these excitations
could move downward in energy and begin to mix with the gapless excitations,
eventually driving a phase transition. Such gapped excitations must be
non-chiral and can only support bosonic excitations. 

A perturbed non-chiral
Luttinger liquid is the simplest example of such a gapped mode:
\begin{multline}
S_{\text{LL}} = \frac{1}{4\pi}\,\int\,dt\,dx\,
\Bigl[ 2 {\partial_t}{\varphi}\,{\partial_x}{\theta} - {v \over g}({\partial_x}\theta)^2 - v g ({\partial_x}\varphi)^2\\
+ {u_1^{(m)}}\cos(m \theta) + {u_2^{(n)}}\cos(n \varphi) \Bigr],
\end{multline}
with Luttinger parameter $g$ and integers $m, n$.
The $\varphi$ and $\theta$ fields have period $2\pi$.
The first line is the action for a gapless Luttinger liquid.
The second line contains perturbations that can open a gap in the Luttinger
liquid spectrum. 
The couplings $u_1^{(m)}$ and $u_2^{(n)}$ have scaling dimensions
$2 - {m^2 \over 2} g$ and $2 - 2 n^2 g^{-1}$, respectively.
Let us concentrate on the lowest harmonics which are the most relevant operators with couplings $u_1^{(1)} \equiv u_1$ and $u_2^{(1)} \equiv u_2$.
The first operator is relevant if $g < 4$ and the second one is relevant
if $g> 1$. At least one of these is always relevant.
Given our parameterization of the Luttinger Lagrangian, a system of hard-core bosons on the lattice with no other interactions
or in the continuum with infinite $\delta$-function repulsion has $g=1$ (see Ref. \onlinecite{Giamarchi2004}). 

When considering one-dimensional bosonic systems, the above cosine perturbations can be forbidden by, respectively,
particle-number conservation and translational invariance. Here, however,
we do not assume that there is any symmetry present, so these terms
are allowed. The Luttinger action can be rewritten in the same way as the edge theory
(\ref{edgetheory}):
\begin{multline}
S_{\text{LL}} = \frac{1}{4\pi}\,\int\,dt\,dx\,
\Bigl[({K_U})_{IJ}{\partial_t}{\phi^I}\,{\partial_x}{\phi^J} 
- V_{IJ} {\partial_t}{\phi^I}\,{\partial_x}{\phi^J}\\
+ {u_1}\cos(\phi_{17}) + {u_2}\cos(\phi_{18}) \Bigr],
\end{multline}
where $I,J=17,18$ in this equation and $\phi_{17} = \theta$ and
$\phi_{18} = \varphi$. Therefore, we see that
the action for a perturbed Luttinger liquid is the edge
theory associated with the trivial bulk theory with $K$-matrix
given by $K_U$ that we discussed in Section \ref{sec:bulk-equivalence}.
It is gapped unless $u_1$ and $u_2$ are fine-tuned to zero or
forbidden by a symmetry. However, augmenting our system with
this trivial one does increase the number of degrees of freedom
at the edge and expands the Hilbert space, unlike in the case of
the bulk.

Hence, we consider the edge theory
\begin{multline}
S = \frac{1}{4\pi}\int \!dt dx\,
\Bigl[{(K_{E_8 \times E_8 \oplus U})_{IJ}}\,{\partial_t}{\phi^I}\,{\partial_x}{\phi^J}
\\ - {V_{IJ}}\,{\partial_x}{\phi^I}\,{\partial_x}{\phi^J}\\
+ {u_1}\cos(\phi_{17}) + {u_2}\cos(\phi_{18})
+ \ldots \Bigr]
\label{edgetheory+U}
\end{multline}
We can integrate out
the trivial gapped degrees of freedom $\phi^{17}$ or $\phi^{18}$, leaving the gapless
chiral edge theory associated with $K_{E_8 \times E_8}$.
The $\ldots$ represents other non-chiral
terms that could appear in the Lagrangian (i.e., cosines
of linear combinations of the fields $\phi^I$);
they are all irrelevant for $V_{I,17}=V_{I,18}=0$
for $I=1,\ldots,16$;
or more accurately, they are less relevant than $u_1$ or $u_2$ and so we ignore them to first approximation. 
However, if we vary the couplings $V_{IJ}$,
then $u_1$, $u_2$ could both become irrelevant and
some other term could become relevant, driving
the edge into another phase.

To further analyze the possible transition, it is useful to rewrite the action
in terms of the fields ${\bf X} = {\bf e}_J \phi^J$:
\begin{multline}
S = \frac{1}{4\pi}\,\int\,dt\,dx\,
\bigl[\eta_{ab}{\partial_t}{X^a}\,{\partial_x}{X^b}
- {v_{ab}}\,{\partial_x}{X^a}\,{\partial_x}{X^b} \\
+ {u_1}\cos(\mbox{$\frac{r}{2}$}(X^{17} + X^{18})) +
{u_2}\cos(\mbox{$\frac{1}{r}$}(X^{17} - X^{18}))
+ \ldots \bigr].
\label{edgetheory2+U}
\end{multline}
where ${v_{ab}}\equiv {V_{IJ}}{f_a^I} {f_b^J}$,
${f_a^I} {e^a_J} = {\bf f}^I \cdot {\bf e}_J =\delta^I_{\ J}$,
and $\eta_{ab} = (1^{16},1,-1)$.
Here, ${\bf e}_J$ for $J=1,\ldots,16$ is a basis
of $\Gamma_{E_8}\oplus\Gamma_{E_8}$ given explicitly in Appendix \ref{sec:lattices-matrices} and $c^n$ refers to the $n$-component vector where each component equals $c$.
We take ${\bf e}_{17} = (0^{16},\frac{1}{r},\frac{1}{r})$ and
${\bf e}_{18} = (0^{16},\frac{r}{2},-\frac{r}{2})$
so that ${\bf e}_{17}\cdot{\bf e}_{17}={\bf e}_{18}\cdot{\bf e}_{18}=0$
and ${\bf e}_{17}\cdot{\bf e}_{18}=1$.
When ${v_{a,17}}={v_{a,18}}=0$ for $a=1,\ldots,16$
(or, equivalently, when $V_{I,17}=V_{I,18}=0$ for $I=1,\ldots,16$),
the parameter $r$ is related to the Luttinger parameter
according to $g={r^2}/2$ and ${u_1}$, ${u_2}$ have renormalization
group (RG) equations:
\begin{eqnarray}
\frac{d{u_1}}{d\ell} &=&
\left(2 - \frac{r^2}{4}\right)\!{u_1},\cr
\frac{d{u_2}}{d\ell}  &=&   \left(2 - r^{-2}\right)\!{u_2}.
\end{eqnarray}
Hence, one of these two perturbations
is always relevant when ${v_{a,17}}={v_{a,18}}=0$ for $a=1,\ldots,16$
and, consequently, $X^{17,18}$ become gapped. 
The arguments of the cosine
follow from the field redefinition $\phi^I = {\bf f}^I \cdot {\bf X} = (K^{-1})^{IJ} {\bf e}_J \cdot {\bf X}$.
The field ${\bf X}$ satisfies the periodicity conditions
${\bf X} \equiv {\bf X} + 2\pi {\bf u}$ for ${\bf u}\in
\Gamma_{E_8}\oplus\Gamma_{E_8} \oplus U$.
Again, the $\ldots$ refers to other possible perturbations,
i.e., cosines of other linear combinations of the $X^a$s.

In a nearly identical manner, we can construct a theory
for ${\rm Spin}(32)/\mathbb{Z}_2 \oplus U$ in which
a non-chiral gapped mode is added to the ${\rm Spin}(32)/\mathbb{Z}_2$
edge theory and allowed to interact with it.
The only difference is in the parameterization of the $U$ lattice. 
We choose ${\bf \tilde{e}}_{17} = (0^{16}, - r, r)$
and ${\bf \tilde{e}}^{18} = (0^{16}, - {1 \over 2r}, - {1 \over 2 r})$.
The action,
\begin{multline}
S = \frac{1}{4\pi}\,\int\,dt\,dx\,
\bigl[\eta_{ab}{\partial_t}{{\tilde X}^a}\,{\partial_x}{{\tilde X}^b}
- {{\tilde v}_{ab}}\,{\partial_x}{{\tilde X}^a}\,{\partial_x}{{\tilde X}^b} \\
+ {{\tilde u}_1}\cos(\mbox{$\frac{1}{2r}$}({\tilde X}^{17} - {\tilde X}^{18})) +
{{\tilde u}_2}\cos(r({\tilde X}^{17} + {\tilde X}^{18}))
+ \ldots \bigr].
\label{edgetheory2+U-alt}
\end{multline}
Again, the $\ldots$ refers to cosines of other linear combinations of the $\tilde{X}^a$s.
When ${\tilde v}_{17,18} = {\tilde v}_{a,17}={\tilde v}_{a,18}=0$ for $a=1,\ldots,16$,
the parameter $r$ is related to the Luttinger parameter
according to $g = r^{-2}/2$ and ${{\tilde u}_1}$, ${{\tilde u}_2}$ have RG equations:
\begin{eqnarray}
\frac{d{{\tilde u}_1}}{d\ell} &=&
\left(2 - \frac{1}{4r^2}\right)\!{{\tilde u}_1},\cr
\frac{d{{\tilde u}_2}}{d\ell}  &=&   \left(2 - r^{2}\right)\!{{\tilde u}_2}.
\end{eqnarray}
Hence, one of these two perturbations
is always most relevant when ${\tilde v}_{a,17}={\tilde v}_{a,18}=0$ for $a=1,\ldots,16$
and, consequently, $X^{17,18}$ become gapped. 
The fields ${\bf {\tilde X}}$ satisfy the periodicity conditions
${\bf {\tilde X}} \equiv {\bf {\tilde X}} + 2\pi {\bf v}$ for ${\bf v}\in
\Gamma_{{\rm Spin}(32)/\mathbb{Z}_2} \oplus U$.

We now make use of the fact there is a unique signature $(17,1)$
even unimodular lattice. It implies that there is an $SO(17,1)$ rotation
$O_G$ that transforms $\Gamma_{E_8}\oplus\Gamma_{E_8} \oplus U$
into $\Gamma_{{\rm Spin}(32)/\mathbb{Z}_2} \oplus U$. Therefore,
the fields ${O_G}{\bf X}$ satisfy the periodicity condition
${O_G}{\bf X} \equiv {O_G}{\bf X} + 2\pi {\bf v}$ for ${\bf v}\in
\Gamma_{{\rm Spin}(32)/\mathbb{Z}_2} \oplus U$
or, in components, $({O_G})^a_{\ b} X^b \equiv ({O_G})^a_{\ b} X^b + 2\pi {n^I}{\tilde e}^a_I$
for $n^I \in \mathbb{Z}$.
Thus, we identify $\tilde{X}^a = (O_G)^a_{\ b} X^b$.
The explicit expression for $O_G$ is provided in Appendix \ref{sec:lattices-matrices}.

(As an aside, having identified $X^a$ and $\tilde{X^b}$ through the $SO(17,1)$ transformation $O_G$, we can now explain how the $SL(18,\mathbb{Z})$ transformation $W_G$ is obtained.
The desired transformation is read off from the relation,
\begin{eqnarray}
\label{linfracbig}
\tilde{\phi}^J = \tilde{f}^{J}_a (O_G)^a_{\ b} e_I^b \phi^I =: (W_G)_{IJ} \phi^I,
\end{eqnarray}
which follows from equation relating the $\Gamma_{E_8} \oplus \Gamma_{E_8}$ and $\Gamma_{{\rm Spin}(32)/\mathbb{Z}_2}$ bases,
\begin{eqnarray}
\label{transformation}
(O_{G})_{\ b}^a e_I^b = \sum_K m^K_I \tilde{e}_K^a,
\end{eqnarray}
where the $m^K_I$ are a collection of integers.
Multiplying both sides of Eq. (\ref{transformation}) by $\tilde{f}^J_c$ allows us to read off the elements of $W_G$.)

Therefore, by substituting $\tilde{X}^a = (O_G)^a_{\ b} X^b$, the action (\ref{edgetheory2+U-alt}) could equally well be
written in the form:
\begin{widetext}
\begin{eqnarray}
S  & =  \frac{1}{4\pi}\int dt\,dx
\bigl[&\eta_{ab} {\partial_t}{X}^a{\partial_x}{X}^b\, 
- {\tilde{v}_{a b}} ({O_G})^a_{\ c} ({O_G})^b_{\ d}
\,{\partial_x}{{X}^c}\,{\partial_x}{X}^d \nonumber
\\ 
& & + {\tilde{u}_1}\cos(\mbox{$\frac{1}{2r}$}(({O_G})^{17}_{\ a} {X}^{a}
- ({O_G})^{18}_{\ a} {X}^{a}))
+ {\tilde{u}_2}\cos(\mbox{$r$}(({O_G})^{17}_{\ a} {X}^{a}
+ ({O_G})^{18}_{\ a} {X}^{a}))
+ \ldots \bigr],
\label{edgetheory2+U-transf}
\end{eqnarray}
\end{widetext}
where ${\bf X} \equiv {\bf X} + 2\pi {\bf u}$ for ${\bf u}\in
\Gamma_{E_8} \oplus \Gamma_{E_8} \oplus U$.
(We have used the defining property, $(O_G)^a_{\ b} \eta_{a c} (O_G)^c_{\ d} = \eta_{b d}$, in rewriting the first term in the action (\ref{edgetheory2+U-alt}).)

Having rewritten the augmented ${\rm Spin}(32)/\mathbb{Z}_2$ action Eq. (\ref{edgetheory2+U-alt}) in terms of the $\Gamma_{E_8} \oplus \Gamma_{E_8}$ fields, let us add in two of the available mass perturbations $u_1, u_2$ written explicitly in Eq. (\ref{edgetheory2+U}):
\begin{widetext}
\begin{eqnarray}
S = \frac{1}{4\pi}\,\int\,dt\,dx\,
\bigl[\eta_{ab}{\partial_t}{X}^a\,{\partial_x}{X}^b - {\tilde{v}_{a b}} ({O_G})^a_{\ c} ({O_G})^b_{\ d}
\,{\partial_x}{{X}^c}\,{\partial_x}{X}^d
+ {\tilde{u}_1}\cos(\mbox{$\frac{1}{2r}$}(({O_G})^{17}_{\ a} {X}^{a}
- ({O_G})^{18}_{\ a} {X}^{a})) \cr
+ {\tilde{u}_2}\cos(\mbox{$r$}(({O_G})^{17}_{\ a} {X}^{a}
+ ({O_G})^{18}_{\ a} {X}^{a}))
+ {u_1}\cos(\mbox{$\frac{r}{2}$}(X^{17}
+ X^{18}))
+ {u_2}\cos(\mbox{$\frac{1}{r}$}(X^{17}
- X^{18}))
+ \ldots \bigr].
\label{edgetheory-combined}
\end{eqnarray}
\end{widetext}
So far we have only rewritten Eq. (\ref{edgetheory2+U-alt}) and included additional mass perturbations implicitly denoted by ``$\ldots$''.
If ${\tilde v}_{17,18}={\tilde v}_{a,17}={\tilde v}_{a,18}=0$ for $a=1,\ldots,16$, then either $\tilde{u}_1$ or $\tilde{u}_2$ is the most relevant operator and the $\tilde{X}^{17}$ and $\tilde{X}^{18}$ fields are gapped out.
The remaining gapless degrees of freedom are those of the ${\rm Spin}(32)/\mathbb{Z}_2$ edge theory.
On the other hand, if ${v}_{c d}=\tilde{v}_{a b} ({O_G})^a_{\ c} ({O_G})^b_{\ d}$ with $v_{17, 18} = v_{a, 17} = v_{a, 18} = 0$, either $u_1$ or $u_2$ is the most relevant operator.
At low energies, $X^{17}$ and $X^{18}$ are gapped with the remaining degrees of freedom being those of the $E_8 \times E_8$ theory. 
We see that the transition between the chiral $E_8 \times E_8$ and ${\rm Spin}(32)/\mathbb{Z}_2$ is mediated by $O_G$ given a starting velocity matrix -- this is an interaction driven transition.


Given $O_G$,
we can define a one-parameter family of $SO(17,1)$ transformations
as follows. As discussed in Appendix \ref{sec:lattices-matrices}, ${O_G}$ can be written
in the form $O_G = \eta W(A) \eta W(A')$, where $W(A), W(A')$ are
$SO(17,1)$ transformations labelled by the vectors $A, A'$ which are defined in Appendix \ref{sec:lattices-matrices} as well and $\eta$ is a reflection. 
We
define ${O_G}(s) = \eta W(sA) \eta W(sA')$.
This family of $SO(17,1)$ transformations, parametrized by $s\in [0,1]$
interpolates between ${O_G}(0)=I$, the identity, and ${O_G}(1)={O_G}$
or, in components, $(O_G(0))^a_{\ b} = \delta^a_{\ b}$, the identity, and
$(O_G(1))^a_{\ b}=(O_G)^a_{\ b}$.
This one-parameter family of
transformations defines a one-parameter family of theories:
\begin{widetext}
\begin{multline}
S_{4}(s) = \frac{1}{4\pi}\,\int\,dt\,dx\,
\bigl[\eta_{ab}{\partial_t}{{X}^a}\,{\partial_x}{X^b} -
{v_{ab}} ({O_G}(s))^a_{\ c} ({O_G}(s))^b_{\ d}
\,{\partial_x}{X^c}\,{\partial_x}{X^d} + {\tilde{u}_1}\cos(\mbox{$\frac{1}{2r}$}(({O_G})^{17}_{\ a} {X}^{a}
- ({O_G})^{18}_{\ a} {X}^{a})) \cr
+ {\tilde{u}_2}\cos(\mbox{$r$}(({O_G})^{17}_{\ a} {X}^{a}
+ ({O_G})^{18}_{\ a} {X}^{a}))
+ {u_1}\cos(\mbox{$\frac{r}{2}$}(X^{17}
+ X^{18}))
+ {u_2}\cos(\mbox{$\frac{1}{r}$}(X^{17}
- X^{18}))
+ \ldots \bigr].\label{edgetheory-family}
\end{multline}
\end{widetext}
These theories are parametrized by $s$, which determines a one-parameter
family of velocity matrices ${v_{ab}} ({O_G}(s))^a_{\ c} ({O_G}(s))^b_{\ d}$
(this is the only place where $s$ enters the action).
We call this action $S_{4}(s)$ because there are $4$ potentially mass-generating
cosine perturbations. Note that the ${\tilde u}_{1,2}$ terms have ${O_G}={O_G}(1)$
in the arguments of the cosines, not ${O_G}(s)$.
As our starting point, we take ${v_{17,18}}={v_{a,17}}={v_{a,18}}=0$ for $a=1,\ldots,16$.
(For instance, we can take diagonal $v_{ab}$.) Then,
for $s=0$, this theory is of the form of Eq. (\ref{edgetheory2+U}) with two extra mass perturbations parameterized by $\tilde{u}_1$ and $\tilde{u}_2$;
however, either ${u}_1$ or ${u}_2$ is most relevant;
and the remaining gapless degrees of freedom
are those of the chiral ${E_8}\times{E_8}$ edge theory.
For $s=1$, this theory is of the form of Eq. (\ref{edgetheory-combined}) which we know is equivalent to Eq. (\ref{edgetheory2+U-alt}) with two extra mass perturbations parameterized by $u_1$ and $u_2$;
now, either ${\tilde{u}}_1$, $\tilde{u}_2$ is most relevant;
and the remaining gapless degrees of freedom
are those of the ${\rm Spin}(32)/\mathbb{Z}_2$ edge theory.
For intermediate values of $s$, the RG equations for
${u}_1$, ${u}_2$, ${\tilde u}_1$, ${\tilde u}_2$ are:
\begin{eqnarray}
\label{interpolatingdims}
\frac{d{u_1}}{d\ell} &=&
\left[2 -\mbox{${(2 s^2 + r^2(1 - s^2 + 4 s^4))^2 \over 4 r^2}$}\right]\!{u_1},\cr
& & \hskip -0.9 cm
\frac{d{u_2}}{d\ell}  =   \left[2 - \mbox{${(1 + 2 r^2 s^2)^2 \over r^2}$}\right]\!{u_2}, \cr
\frac{d{{\tilde u}_1}}{d\ell} &=&
\left[2 -  \mbox{${(4 - 7 s + 4 s^2 + 2 r^2(s-1)^2(1 + s + 4 s^2))^2 \over 4 r^2}$}\right]
\!{{\tilde u}_1}, \cr
& & \hskip -0.9 cm \frac{d{{\tilde u}_2}}{d\ell}  =  
\left[2 - \mbox{${(2 (s - 1)^2 + r^2 (1 + s + 3 s^2 - 8 s^3 + 4 s^4))^2 \over r^2}$}\right]
\!{{\tilde u}_2}.
\end{eqnarray}
The expressions in square brackets on the right-hand-sides of these equations,
which are equal to ${1 \over u_{1,2}}\frac{du_{1,2}}{d\ell}$
and ${1 \over \tilde{u}_{1,2}}\frac{d{\tilde u}_{1,2}}{d\ell}$, are the scaling dimensions
of $u_{1,2}$ and ${\tilde u}_{1,2}$ near the $u_{1,2}={\tilde u}_{1,2}=0$ fixed line.

\begin{figure}[t]
\begin{center}
\includegraphics[width=0.45\textwidth]{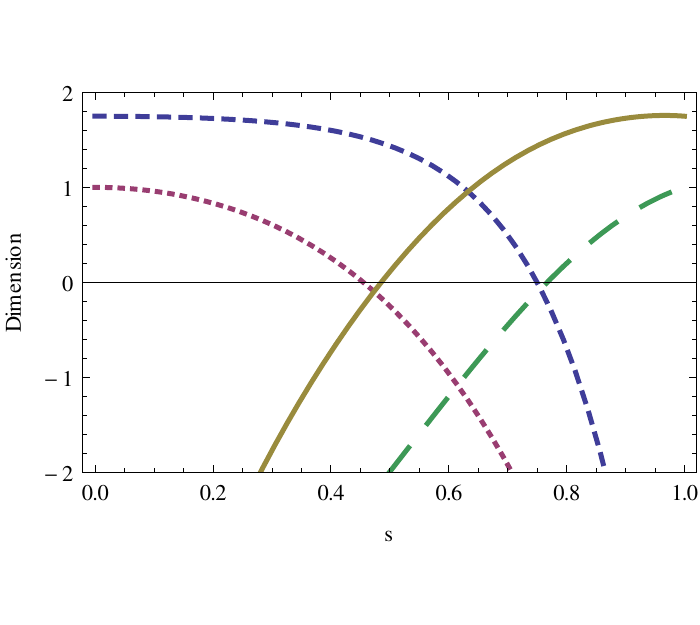}
\end{center}
\caption{The scaling dimensions of $u_{1,2}$ (densely dashed and dotted) and ${\tilde u}_{1,2}$ (thick and dashed),
plotted as a function of $s$ at $r=1$.
The $E_8 \times E_8$ phase lives roughly within $0 \leq s < .625$ and the ${\rm Spin}(32)/\mathbb{Z}_2$ phase between $.625 < s \leq 1$.}
\label{figone}
\end{figure}

\begin{figure}[t]
\begin{center}
\includegraphics[width=0.45\textwidth]{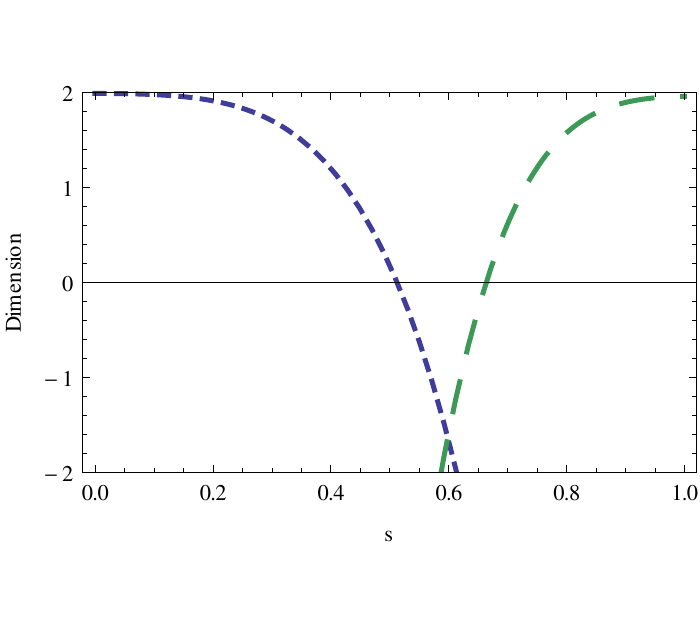}
\end{center}
\caption{
The scaling dimensions of $u_{1}$ (densely dashed) and ${\tilde u}_{2}$ (dashed),
plotted as a function of $s$ at  at $r=.2$.
The scaling dimensions of $u_{2}$ and ${\tilde u}_{1}$
lie outside the range of the plot and are not displayed.
The system is not fully chiral phase between approximately $s = .5$ and $s = .625$.}
\label{figtwo}
\end{figure}

\begin{figure}[t]
\begin{center}
\includegraphics[width=0.45\textwidth]{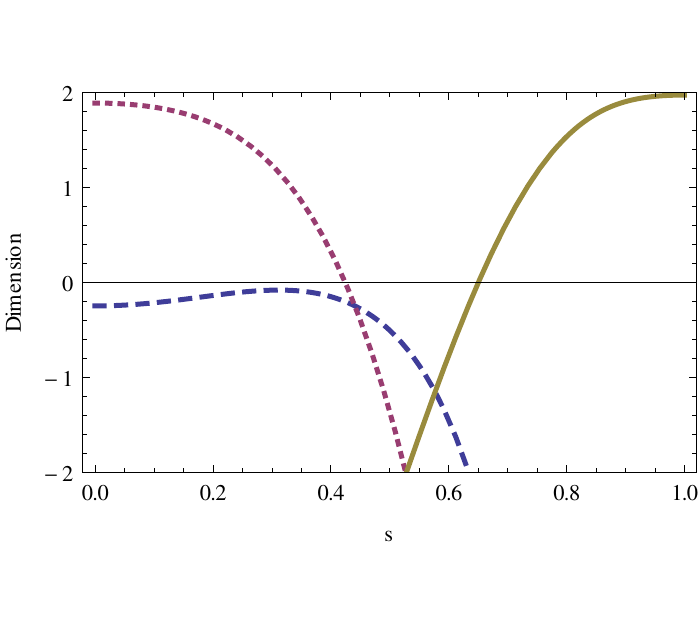}
\end{center}
\caption{The scaling dimensions of $u_{1,2}$ (densely dashed and dotted)
and ${\tilde u}_{1}$ (thick),
plotted as a function of $s$ at  at $r=3$.
The scaling dimension of ${\tilde u}_{2}$
lies outside the range of the plot and is not displayed.
The system is in the not fully chiral phase between
approximately $s = .425$ and $s = .625$.}
\label{figthree}
\end{figure}

We plot the weak-coupling RG flows of these operators in
Figs. \ref{figone}-\ref{figthree}
for three different choices of $r$.
First, we notice that, depending upon $r$, either $u_1$ or $u_2$
is most relevant at $s=0$. 
At $s=1$, either $\tilde{u}_1$ or $\tilde{u}_2$ is most relevant.
At intermediate values of $s$, there are several possibilities.
Assuming that the most relevant operator determines the flow
to low energy (which must have the same value $c-\overline{c}=16$
as the action (\ref{edgetheory-family})), we conclude that
when either of these two sets of operators is most relevant we expect a mass to be generated for, respectively, the $X^{17,18}$ or $\tilde{X}^{17,18}$ modes,
thereby leaving behind either the $E_8 \times E_8$ or ${\rm Spin}(32)/\mathbb{Z}_2$
edge theories at low energies. If there are no relevant operators, then
the edge is not fully chiral; it has $c=17$, $\overline{c}=1$.


Thus, we see that the two different positive-definite even unimodular lattices
in $16$ dimensions correspond to two different fully chiral phases at the edge
of the same bulk phase. In the model in Eq. (\ref{edgetheory-family}),
the transition between them can occur in two
possible ways: either a direct transition (naively, first-order, as we argue below)
or or via two Kosterlitz-Thouless-like phase transitions, with
an intermediate $c=17$, $\overline{c}=1$ phase between the two
fully chiral phases. The former possibility occurs
(again, assuming that the most relevant operator determines the flow
to low energy) when there is always at least one relevant operator.
The system is in the minimum of the corresponding cosine, but
when another operator becomes more relevant, the system jumps
to this minimum as $s$ is tuned through the crossing point.
Precisely at the point where two operators are equally-relevant
(e.g. $u_1$ and ${\tilde u}_1$ at $r=1, s\approx 0.6$ as shown in
Fig.~\ref{figone})
the magnitudes of the two couplings become important. At a mean-field
level, the system will be in the minimum determined by the larger coupling
and there will be a first-order phase transition at the point at which these
two couplings are even in magnitude.

If the most relevant operator is in the set ${u_1}, {u_2}, {{\tilde u}_1}, {{\tilde u}_2}$,
then this means that the crossing point between the larger of
$\frac{1}{u_{1,2}}\frac{du_{1,2}}{d\ell}$ and the larger of
$\frac{1}{{\tilde u}_{1,2}}\frac{d{\tilde u}_{1,2}}{d\ell}$
occurs when both are positive so that the system goes directly from
$E_8 \times E_8$ to ${\rm Spin}(32)/\mathbb{Z}_2$ theory.
However, if there is a regime in which there are no relevant operators, then there will
be a stable $c=17$, $\overline{c}=1$ phase. 
(Note that we adhere to a slightly weaker definition of stability than used in the recent paper \cite{Levin13}; we say that an edge is unstable to gapping out some subset of its modes if a null vector \cite{Haldanestability} of the $K$-matrix exists and that the associated operator is relevant in the RG sense.
A null vector is simply an integer vector $n_I$ satisfying $n_I (K^{-1})^{IJ} n_J = 0$ or, equivalently, a lattice vector $k_a$ satisfying $k_a \eta^{a b} k_b = 0$.)
If the crossing point
between the larger of $\frac{1}{u_{1,2}}\frac{du_{1,2}}{d\ell}$ and the larger of
$\frac{1}{{\tilde u}_{1,2}}\frac{d{\tilde u}_{1,2}}{d\ell}$
occurs when both are negative, then there may be a stable
$c=17$, $\overline{c}=1$ phase.

\begin{figure}[t]
\begin{center}
\includegraphics[width=0.4\textwidth]{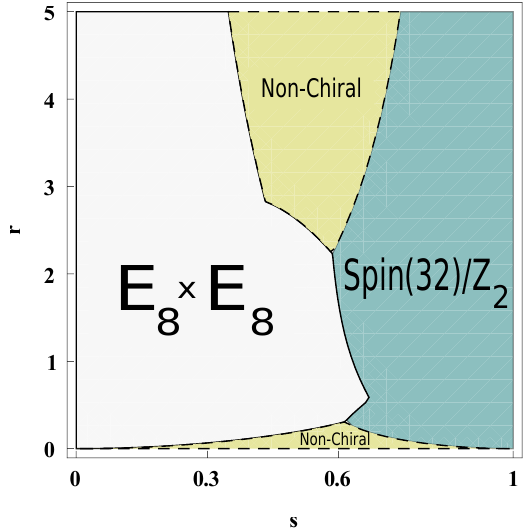}
\end{center}
\caption{Phase diagram of our edge theory as a function of $s$ and $r$
for the theory ${S_4}(s)$ in which the only non-zero
perturbations are ${u_{1,2}}$ and ${\tilde u}_{1,2}$. 
The light region is in the $E_8 \times E_8$ phase. 
The darkest region is in the ${\rm Spin}(32)/\mathbb{Z}_2$ phase. 
The system is not fully chiral in the intermediately-shaded region.
The dashed phase boundary line indicates a KT transition. The solid
lines denote regions where there are two equally-relevant couplings;
the phase is determined by their ratio.}
\label{figfour}
\end{figure}

However, the model of Eq. (\ref{edgetheory-family}) is not the most
general possible model; it is a particular slice of the parameter
space in which the only perturbations of the quadratic theory
are ${u_{1,2}}$ and ${\tilde u}_{1,2}$. A more general
model will have many potentially mass-generating perturbations:
\begin{equation}
\label{eqn:dangerous-operators}
S_{\text{gen}}(s) = S_{4}(s) + \int\,dt\,dx\, \!\!\!\!\!\sum_{\scriptscriptstyle {{\bf v} \in \Gamma_{E_8} \oplus \Gamma_{E_8} \oplus U}}
\delta_{{|{\bf v}|^2},0}\, u_{{\bf v},s}\cos({{\bf v}\cdot {\bf X}})
\end{equation}
where the sum is over vectors ${\bf v} \in \Gamma_{E_8} \oplus \Gamma_{E_8} \oplus U$
that have zero norm. This guarantees that these are spin-$0$ operators
that are mass-generating if relevant. In Eq. (\ref{edgetheory-family}), we have chosen
$4$ particular operators of this form and set the coefficients of the others to zero.
\footnote{To lowest-order in ${u_{1,2}}$ and ${\tilde u}_{1,2}$, this is consistent, but at
higher order, these $4$ operators will generate some others, and we must consider
a more general theory. However, it does not appear that these operators
generate any spin-$0$ operators other than multiples of themselves,
which are less relevant than they are.}
However, to determine if there is a stable non-chiral phase,
it behooves us to consider a more
general model in order to determine whether the non-chiral phase requires
us to set more than one of the potentially mass-generating operators
in Eq. (\ref{eqn:dangerous-operators}) to zero by hand and so any such critical point is multi-critical.

Of course, there are many possible
${\bf v} \in \Gamma_{E_8} \oplus \Gamma_{E_8} \oplus U$
with ${|{\bf v}|^2} = 0$. But most of them give rise to operators that are highly irrelevant
over most of the range of the parameters $r$ and $s$. However, there
are two sets of operators that cannot be ignored. In one set, each operator
is highly relevant in the vicinity of a particular value of $s$ (which depends
on the operator) in the $r\rightarrow 0$ limit and, in the other set, each operator
is highly relevant in the vicinity of a particular value of $s$
in the $r\rightarrow \infty$ limit. Consider the operators:
\begin{equation}
\cos(\alpha {\tilde f}^{17}_{\ a} R^a_{\ b} {X^b}) \,\, ,
\hskip 0.2 cm \cos(\beta {\tilde f}^{18}_{\ a} R^a_{\ b} {X^b}) 
\end{equation}
where $R$ is an arbitrary $SO(17,1)$ transformation.
These operators have spin-$0$ since $\tilde{{\bf f}}^{17,18}$ have vanishing norm,
which $R$ preserves. Although they have spin-$0$ and can, therefore,
generate a mass gap, there is no particular reason to think that either one
is relevant. Moreover, it is not even likely that either one is an allowed operator.
For an arbitrary $SO(17,1)$ transformation, $\tilde{f}^{17}_{\ a} R^a_{\ b}$ will not
lie in the $\Gamma_{E_8} \oplus \Gamma_{E_8} \oplus U$ lattice spanned
by the $f^I$s, so this operator will not be allowed. However, there is a special class of $R$
for which these operators are allowed and are relevant in the vicinity of special
points. Let us suppose that $R={O_G}(p/q)$ and let us consider
$\alpha=q^4$, $\beta = q^2$.
\footnote{This choice of $\alpha$ and $\beta$ is a sufficient one for generic $s = p/q$; however, certain $q$ accommodate smaller $\alpha$ and $\beta$ so that the resulting operators are well defined. 
For example, when $q$ is even, we may take $\alpha = q^2/2$ and $\beta = q^4/4$.}
Consider the action
\begin{equation}
S_{4}(s=\mbox{$\frac{p}{q}$}) + u_{18,{\scriptscriptstyle \frac{p}{q}}}\int\,dt\,dx\,\cos\left[{q^2} {\tilde f}^{18}_{\ a} \left({O_G}(p/q)\right)^a_{\ b} {X^b}\right]
\end{equation}
This is a spin-$0$ perturbation.
Moreover, it is an allowed operator for the following reason. We can write
\begin{equation}
{q^2} {\tilde f}^{18}_{\ a} \left({O_G}(p/q)\right)^a_{\ b} = {q^2}(W(p/q))_{18, J} f^J_{\ a}
\end{equation}
where $(W(s))_{IJ}$ is defined in analogy with $W_G$:
$(W(s))_{IJ} = \tilde{f}^{J}_a ({O_G}(s))^a_{\ b} e_I^b$.
The vector ${q^2}(W(p/q))_{18, J}$ has integer entries,
so  ${q^2} {\tilde f}^{18}_{\ a} \left({O_G}(p/q)\right)^a_{\ b}$
is in the lattice $\Gamma_{E_8} \oplus \Gamma_{E_8} \oplus U$. 
At the point $s=p/q$, its scaling dimension
is the same as the scaling dimension of ${q^2}{\tilde f}^{18}_{\ a}{X^a}$ at $s=0$:
\begin{equation}
\frac{d{}}{d\ell} u_{18,{\scriptscriptstyle \frac{p}{q}}} =
\left[2 - {q^4} r^2\right]u_{18,{\scriptscriptstyle \frac{p}{q}}}
\end{equation}
Therefore, for
$r<\sqrt{2}/q^2$, the coupling $u_{18,{\scriptscriptstyle \frac{p}{q}}}$
is a relevant mass-generating interaction at $s=p/q$ and, over some range
of small $r$, it is relevant for $s$ sufficiently near $p/q$. By a similar analysis,
$u_{17,{\scriptscriptstyle \frac{p}{q}}}$ is a relevant mass-generating interaction
at $s=p/q$ for $r>q^4/(2\sqrt{2})$ and, over some range of large $r$,
it is relevant for $s$ sufficiently near $p/q$. Therefore, when these couplings are
non-zero, the non-chiral phase survives in a much smaller region of the phase diagram.
(Making contact with our previous notation, we see that $u_{17,1}= \tilde{u}_1$
and $u_{18,1} = \tilde{u}_2$.)

When one of these interactions gaps out a pair of counter-propagating modes,
we are left with a fully chiral $c=16$ edge theory corresponding to either 
$E_8 \times E_8$ to ${\rm Spin}(32)/\mathbb{Z}_2$. To see which phase
we get, consider, for the sake of concreteness, the coupling
$u_{18,{\scriptscriptstyle \frac{p}{q}}}$. When it generates a gap, it
locks the combination of fields
${q^2}{\tilde f}^{18}_{\ a} \left({O_G}(p/q)\right)^a_{\ b} {X^b}=
{q^2}(W(p/q))_{18, J} f^{J}_{\ a}{X^a}$.
In the low-energy limit, we may set this combination to zero.
Only fields that commute with this combination remain gapless.
(Moreover, since we have set this combination to zero, any fields that
differ by a multiple of it are equal to each other at low-energy.)
Therefore, the vertex operators that remain in the theory are of
the form $\exp({n_I}f^J_{\ a}{X^a})$ where ${n_I}$ satisfies
${n_I}(K^{-1})^{IJ}(W(p/q))_{18, J}=0$. We note that $(W(p/q))_{18,J}$
is non-zero only for $J=8,16,17,18$. Therefore, $(W(p/q))_{18,J}f^J_{\ a}$
is orthogonal to ${{\bf e}_1},\ldots,{{\bf e}_7}$ and ${{\bf e}_9},\ldots,{\bf e}_{15}$.

Much as in our discussion in Section \ref{sec:fermionic} of the difference between the $E_8 \times E_8$ and ${\rm Spin}(32)/\mathbb{Z}_2$ edge theories, we again make use of the basic observation that $E_8 \times E_8$ is a product while ${\rm Spin}(32)/\mathbb{Z}_2$ has a single component in order to identify the low energy theory.
If the vectors ${n_I}f^I_{\ a}$ with ${n_I}(K^{-1})^{IJ}(W(p/q))_{18,J}=0$
(and two vectors differing by a multiple of ${q^2}(W(p/q))_{18,J} f^J_{\ a}$
identified) form the ${\rm Spin(32)}/\mathbb{Z}_2$ lattice, then there
must be a vector ${\bf c} = {c_I}{\bf f}^{I}$ in the lattice with $|{\bf c}|^2 = 2$
such that ${\bf c}\cdot {{\bf e}_1} = - {\bf c}\cdot {{\bf e}_7} = {\bf c}\cdot {{\bf e}_9}=1$
and ${\bf c}\cdot {{\bf e}_2} = {\bf c}\cdot {{\bf e}_3} = \ldots = {\bf c}\cdot {{\bf e}_6}=0$
and ${\bf c}\cdot e_{10} = {\bf c}\cdot e_{11} = \ldots = {\bf c}\cdot e_{15}=0$.
This is because there exists a set of Cartesian coordinates ${\bf \hat y}_a$
such that all the vectors in ${\rm Spin(32)}/\mathbb{Z}_2$ with $(\text{length})^2=2$
are of the form $\pm{\bf \hat y}_a \pm {\bf \hat y}_b$ with $a,b=1,\ldots,16$,
while for $E_8 \times E_8$, vectors of the form $\pm{\bf \hat y}_a \pm {\bf \hat y}_b$
must have $a,b=1,\ldots, 8$ or $a,b=9,\ldots, 16$. In $E_8 \times E_8$, vectors
of $(\text{length})^2=2$ cannot ``connect'' the two halves of the system.
If the equations ${c_I}(K^{-1})^{IJ}(W(p/q))_{18,J}=0$  and ${c_I}(K^{-1})^{IJ}c_{J}=2$
with $c_1 = -c_7 = c_9 = 1$ and
$c_2 = c_3 = \ldots = c_6 = c_{10} = c_{11} = \ldots = c_{15}=0$
have integer solutions, then the remaining gapless degrees of freedom
are in the ${\rm Spin(32)}/\mathbb{Z}_2$ phase. Otherwise, they are in the
$E_8 \times E_8$ phase. 
We could choose ${\bf e}_1$, $-{\bf e}_7$,
and ${\bf e}_9$ as the vectors with unit product with ${\bf c}$ because
such a ${\bf c}$ must exist in ${\rm Spin(32)}/\mathbb{Z}_2$.
(Note, that we could have taken $c_7$ to be arbitrary, and we would
have found that solutions to these equations must necessarily
have ${c_7}=-1$.)
The phase is $E_8 \times E_8$ if and only if such a vector ${\bf c}$ is not in the lattice. 
Of course, it is essential that we can restrict our attention to the two
possibilities, $E_8 \times E_8$ and ${\rm Spin(32)}/\mathbb{Z}_2$,
since these are the only two unimodular self-dual lattices in dimension 16.

With the aid of {\sc Mathematica}, we have found that
solutions to the above equations must be of the form
$c_I = (1, 0^5, -1, c_8, 1, 0^6, c_8 - 1, q/p(2 c_8 - 1), - p/q(2 c_8 - 1))$.
Since $c_I$ must be an integer vector, both $p$ and $q$ must be
odd since $2 c_8 - 1$ is odd. 
Here, as above, we have assumed that $p$ and $q$ are relatively prime.   
Further, we see that this solution requires $2 c_8 = p q m + 1$ for odd $m$.

This means that the chiral ${\rm Spin}(32)/\mathbb{Z}_2$ theory is left behind at low energies when both $p$ and $q$ are odd and $u_{18, {\scriptscriptstyle \frac{p}{q}}}$ is the most relevant
operator that generates a mass gap for two counter-propagating edge modes.
When either $p$ or $q$ is even, the remaining gapless modes of the edge are in the $E_8 \times E_8$ phase.
We find the identical behavior for the low energy theory when $u_{17,{\scriptscriptstyle \frac{p}{q}}}$ is the most relevant operator.

When these operators have non-zero coefficients in the Lagrangian,
they eliminate a great deal of the non-chiral phase shown in the
$u_{1,2}$, ${\tilde u}_{1,2}$-only phase diagram in
Fig. 4. 
The effect is most noticeable as $r \rightarrow 0$ and $r \rightarrow \infty$ as shown in Fig. 5.

However, there still remain pockets of the non-chiral phase
at intermediate values of $r$ and $s$, where these operators are irrelevant.
However, we find that these regions of non-chiral phase
are not stable when we include a larger set of operators in the Lagrangian.
Consistent with our expectations, it is possible to find a relevant operator in the region around any given point $(r,s)$ in the phase diagram such that the low energy theory remaining after a pair of counter-propagating modes gaps out is $E_8 \times E_8$ or ${\rm Spin}(32)/\mathbb{Z}_2$.

To see how this works, consider, for instance, the point $(r, s) = (3, 3/5)$ that exists in the putative region of non-chiral phase according to Fig. 4. The couplings
$u_{17,{\scriptscriptstyle \frac{p}{q}}}$, $u_{18,{\scriptscriptstyle \frac{p}{q}}}$
are all irrelevant there so the system remains non-chiral even when these
couplings are turned on. However, we can find a relevant spin-0 operator
at this point as follows. It must take the form
$\cos(p_a X^a)$, with ${p_a}\in {\Gamma_8}\oplus{\Gamma_8}\oplus U$,
where $\eta^{ab}{p_a}{p_b}=0$ (this is the spin-$0$ condition).
To compute its scaling dimension, we observe that it can be written
in the form $\cos( q_a {X^a}(s))$, where ${X^a}(s) \equiv (O_G(s))^a_b X^b$
and $p_b = q_a (O_G(s))^a_b$.
In terms of this field, the quadratic part of the action is diagonal
in the ${X^a}(s)$ fields, so their correlation functions (and, therefore,
their scaling dimensions can be computed straightforwardly).
Since the operator in question has spin-$0$, its total scaling
dimension $\delta^{ab}{q_a}{q_b}$
is twice their left-moving dimension or, simply, $|q_{18}|^2$.
Therefore, such an operator is relevant if $|q_{18}|^2 < 2$.

$O_G^{-1}(s)$ is simply a boost along some particular direction in the 17-dimensional space combined with a spatial rotation. The eigenvalues of such a transformation are
either complex numbers of modulus $1$ (rotation) or contraction/dilation
by $e^{\pm \alpha}$ (Lorentz boost). Consequently, even if
$\delta^{ab}{p_a}{p_b}$ is large -- which means that $\cos( p_a X^a)$
is highly irrelevant at $s=0$ -- $\delta^{ab}{q_a}{q_b}$ can be smaller by
as much as $e^{-2\alpha}$, thereby making $\cos( p_a X^a)$ a relevant
operator at this value of $s$ (and of $r$). The maximum possible
contraction, $e^{-\alpha}$, occurs when $p_a$ is anti-parallel to
the boost. (The maximum dilation, $e^{-\alpha}$, occurs when $p_a$
is parallel to the boost, and there is no change in the scaling dimension
when $p_a$ is perpendicular to the boost.) For a given $r, s$,
we can choose a lattice vector $p_a$ that is arbitrarily close
to the direction of the boost, but at the cost of making $\delta^{ab}{p_a}{p_b}$
very large. Then $\delta^{ab}{q_a}{q_b}\approx e^{-2\alpha}\delta^{ab}{p_a}{p_b}$
may not be sufficiently small to be relevant. (The $\approx$ will be an
$=$ sign if $p_a$ is precisely parallel to the direction of the boost, however, we are not guaranteed to be able to find an element of the lattice that is precisely parallel.)
Alternatively, we can choose a smaller $\delta^{ab}{p_a}{p_b}$,
but the angle between $p_a$
and the boost may not larger. As explained through an example
in Appendix \ref{sec:dimension-contraction}, we can balance these
two competing imperatives and find a $p_a$ so that neither
$\delta^{ab}{p_a}{p_b}$ nor the angle between $p_a$
and the boost is too large. Then
$\frac{1}{2}\delta^{ab}{q_a}{q_b}\approx \frac{1}{2}e^{-2\alpha}\delta^{ab}{p_a}{p_b}<2$,
so that the corresponding operator is relevant. 

The following simple ansatz
leads to a relevant operator 
\begin{equation}
p_a = n f^7_{\ a} + (m - 2 n) f^8_{\ a} + m f^{16}_{\ a} + n_{17} f^{17}_{\ a} + n_{18} f^{18}_{\ a}
\end{equation}
at all candidate non-chiral points in the $(r,s)$ phase diagram
that we have checked. We do not have a proof that there is not
some region in parameter space where a non-chiral phase is stable,
but we have explicitly excluded nearly all of it, as may be seen from
the phase diagram in Fig. \ref{figsix} where we have included a selection of the possible operators described here that become relevant at the set of points $(r,s) = (6, p/q)$ for $q = 5$, and we anticipate that
this ansatz will enable us to do so for any other point not already
excluded. Thus, we expect the non-chiral phase to be entirely
removed by this collection of operators combined with those discussed earlier.

Therefore, the phase diagram has a quite rich and intricate structure. 
From our experience with the above operators, our general expectation is that in the neighborhood of any point $(r, p/q)$, there exists a relevant operator that gaps out a pair of modes leading to the fully chiral $E_8 \times E_8$ theory if $p$ or $q$ is even, while ${\rm Spin}(32)/\mathbb{Z}_2$ remains if $p$ and $q$ are odd.

\begin{figure}[t]
\begin{center}
\end{center}
\includegraphics[width=0.21\textwidth]{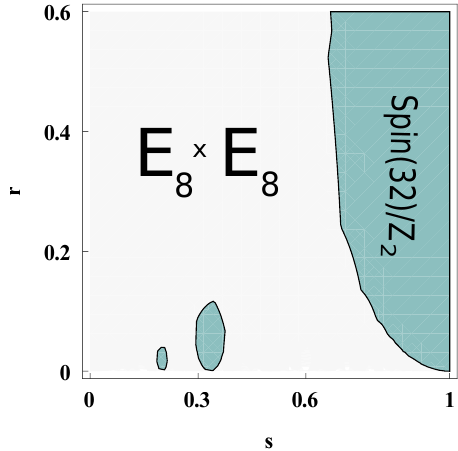}
\includegraphics[width=0.22\textwidth]{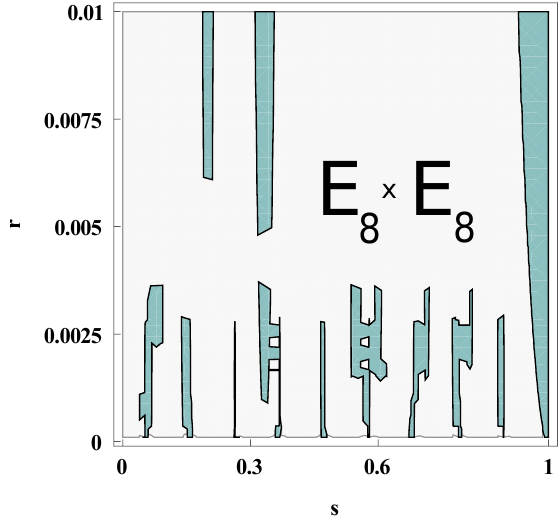}
\caption{The small-$r$ region of the phase diagram of our
edge theory as a function of $s$ and $r$
for the theory with non-zero ${u_{1,2}}$, ${\tilde u}_{1,2}$;
$u_{17,{\scriptscriptstyle \frac{p}{q}}}$, $u_{17,{\scriptscriptstyle \frac{p}{q}}}$
for all $p,q\leq 57$; and several $\cos(p_a X^a)$ operators with
$p_a$ nearly aligned with the direction of the boost $O_G(s)$,
as described in the text. 
The light region is in the $E_8 \times E_8$ phase. 
The darker region is in the ${\rm Spin}(32)/\mathbb{Z}_2$ phase. 
All phase boundary lines denote regions where there are two equally-relevant
couplings; the phase is determined by the ratio of these couplings.
The left panel shows the $r<0.6$ region of the phase diagram,
where we see that regions of the two
phases are interspersed with each other along the $s$-axis.
In the right panel, we zoom in on the $r<0.01$ region of the phase diagram
and see an even richer intermingling of these
two phases as we sweep over $s$.}
\label{figfive}
\end{figure}

\begin{figure}[t]
\begin{center}
\includegraphics[width=0.4\textwidth]{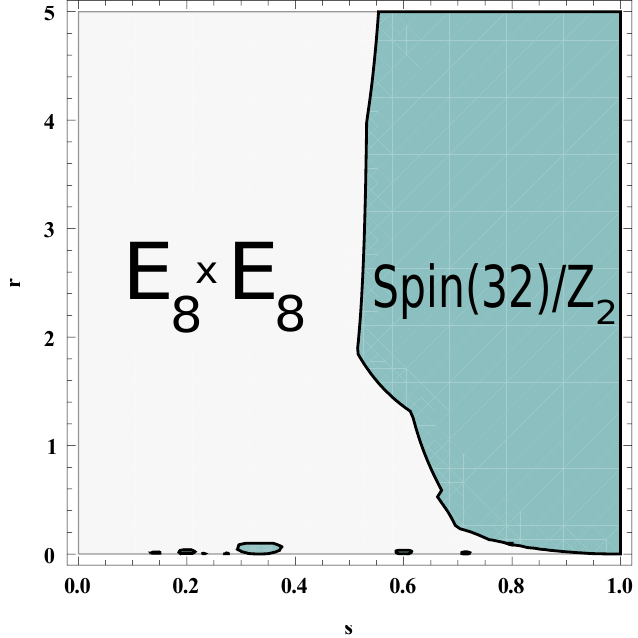}
\end{center}
\caption{Phase diagram of our edge theory as a function of $s$ and $r$
for the theory ${S_4}(s)$ in which the only non-zero
perturbations are ${u_{1,2}}$ and ${\tilde u}_{1,2}$;
$u_{17,{\scriptscriptstyle \frac{p}{q}}}$, $u_{17,{\scriptscriptstyle \frac{p}{q}}}$
for all $p,q\leq 57$; and several $\cos(p_a X^a)$ operators with
$p_a$ nearly aligned with the direction of the boost $O_G(s)$,
as described in the text.
The latter operators were specifically chosen to remove the remaining points of non-chiral phase at $r=6$, $s=p/q$ for $q = 5$. 
This set of operators was sufficient to remove all the non-chiral phase displayed previously in Fig. 4. 
The light region is in the $E_8 \times E_8$ phase. 
The darker region is in the ${\rm Spin}(32)/\mathbb{Z}_2$ phase. 
Solid phase boundary lines denote regions
where there are two equally-relevant
couplings; the phase is determined by the ratio of these couplings.}
\label{figsix}
\end{figure}

\section{Charged Systems}
\label{sec:t-vectors}

We return to our $c-\overline{c}=16$ theories and consider the
case in which some of the degrees of freedom are charged
as a result of coupling to an external electromagnetic field
as in Eq. (\ref{cstheory-with-t}).
Now, there are many phases for a given $K$,
distinguished by different $t$. They may,
as a consequence, have different Hall conductances
$\sigma_{xy} = \frac{e^2}{h} t_I (K^{-1})^{IJ} t_J$,
which must be even integer multiples of $\frac{e^2}{h}$
since $K^{-1}$ is an integer matrix with even entries on the diagonal.

Let us focus on the minimal possible non-zero Hall conductance,
$\sigma_{xy} = 2 \frac{e^2}{h}$. We will not attempt to systematically
catalog all of these states here, but will examine a few examples 
with $c=16$ that are enlightening. By inspection, we see that
we have three distinct $\sigma_{xy} = 2 \frac{e^2}{h}$ states with
$K$-matrix $K=K_{E_8 \times E_8}$: (1) $t_I = \delta_{I6}$,
(2) $t_I = \delta_{I9}$, and (3) $t_I = -2\delta_{I1}+\delta_{I2}$.
These states have stable edge modes even if
the $U(1)$ symmetry of charge conservation is violated
(e.g., by coupling the system to a superconductor), in contrast
to the $\sigma_{xy} = 2 \frac{e^2}{h}$ bosonic quantum Hall states
discussed in Ref. \onlinecite{Senthil13}.

As before, we adjoin a trivial system to our system so that
the $K$-matrices are $K=K_{E_8 \times E_8 \oplus U}$.
Under the similarity transformation $W_G$, these states
are equivalent to the states with $K_{\text{Spin}(32)/\mathbb{Z}_2 \oplus U}$
and, respectively, $t=(0, 0, 0, 0, 1, -2, 0, 0, 0, 0, 0, 0, 0, 0, 0, 4, -2, 2)$,
$t={0, 0, 0, 0, 0, 0, 0, -2, 1, 0, 0, 0, 0, 0, 0, 4, -2, 2}$,
and $t_I=\delta_{I1}$. Consider the first of these,
$K_{\text{Spin}(32)/\mathbb{Z}_2 \oplus U}$,
$t=(0, 0, 0, 0, 1, -2, 0, 0, 0, 0, 0, 0, 0, 0, 0, 4, -2, 2)$.
It is {\it not} equal to $K_{\text{Spin}(32)/\mathbb{Z}_2}$ with
an additional trivial system adjoined to it because ${\tilde \phi}_{17}$
and ${\tilde \phi}_{18}$ are both charged. In other words,
there is a right-moving neutral edge mode ${\tilde \phi}_{17}+{\tilde \phi}_{18}$
and a left-moving charged edge mode ${\tilde \phi}_{17}-{\tilde \phi}_{18}$.
This is non-trivial, and there is no charge-conserving perturbation which
will give a gap to these modes. The same is true of the second state.
In the case of the third state, both ${\phi}_{17}$, ${\phi}_{18}$
and ${\tilde \phi}_{17}$, ${\tilde \phi}_{18}$ are neutral. Therefore,
there are perturbations that could gap out either of them.
Consequently, we conclude that $K=K_{E_8 \times E_8}$,
$t_I = -2\delta_{I1}+\delta_{I2}$ and
$K_{\text{Spin}(32)/\mathbb{Z}_2}$, $t_I=\delta_{I1}$
are stably equivalent bulk states with a edge theory phase diagram
similar to that in Figure \ref{figfour}.

\section{Discussion}

\label{sec:conclusion}

\subsection{Summary}

Bosonic SRE states with chiral edge modes are bosonic
analogues of fermionic integer quantum Hall states:
they do not support anyons in the bulk, but they have
completely stable chiral edge modes.
Together, they populate an `intermediate' class of phases that are completely
stable and do not require symmetry-protection, however, they lack non-trivial
bulk excitations.
Unlike in the fermionic
case, such states can only occur when the number of edge modes
is a multiple of $8$.
As we have seen in this paper, the scary possibility that
the number of edge modes does not uniquely determine
such a state is not realized, at least for the first case in which
it can happen, namely, when there are $16$ edge modes.
The two phases that are naively different are, in fact, the same phase.
This is consistent with the result that all $3$-manifold invariants
associated with the two phases are the same \cite{Belov05},
and we have gone further and shown that it is possible to go directly
from one state to the other
without crossing a phase boundary in the bulk. However, there are actually
two distinct sets of edge excitations corresponding to these adiabatically
connected bulk states. We have shown that the phase transition
between them can occur purely at the edge, without closing the bulk gap.
However, both edge phases are fully chiral, unlike the ``$T$-unstable''
states considered in Refs. \onlinecite{Haldanestability,Kao99}.
There is no sense in which one of these two phases is inherently more
stable in a topological sense than the other; it is simply that, for some values
of the couplings, one or the other is more stable.

Our construction is motivated by the observation that there is a unique
even, unimodular lattices with signature $(8k+n,n)$. Consequently, enlarging
the Hilbert spaces of seemingly different phases associated with distinct
even, unimodular lattices with signature $(8k,0)$ by adding trivial insulating
degrees of freedom associated with even, unimodular lattices with
signature $(n,n)$ leads to the same bulk phase. Since the edge is characterized
by additional data, the corresponding edge theories are distinct but are separated
by a phase transition that can occur purely on the edge without closing the bulk
gap. The details of our construction draw on a similar one by Ginsparg
\cite{Ginsparg87} who showed explicitly how to interpolate
between toroidal compactifications of ${E_8} \times {E_8}$
and $\text{Spin}(32)/\mathbb{Z}_2$ heterotic string theories.

\subsection{Future Directions}

\label{sec:future}

Let us describe a few possible directions for future study. 

\begin{itemize}

\item We have considered one possible interpolation between the $E_8 \times E_8$ and ${\rm Spin(32)}/\mathbb{Z}_2$ theories and, therefore, have only considered a small region of possible parameter space determined by $r$ and $s$. 
It would be interesting to carve out in more detail the full 153-dimensional phase space.

\item The last phase diagram displayed in Fig. \ref{figsix} includes only a subset of the possible operators that may be added to the edge theory.
The operators that have been added are sufficient to lift the non-chiral phase that is naively present and displayed in Fig. \ref{figfour} when only four operators are included.
It is possible that consideration of all allowed operators could result in an
even more complex phase diagram with a rich topography
of interspersed $E_8 \times E_8$ and ${\rm Spin}(32)/\mathbb{Z}_2$ phases.

\item The uniqueness of even, unimodular lattices with signature
$(8k+n,n)$ implies that a similar route can be taken to adiabatically
connect states associated to different positive-definite even unimodular
lattices of dimension $8k=24,32,\ldots$. However, in these cases,
it is possible for states corresponding to different lattices
to have different spectra of operator scaling dimensions
at the edge, unlike in the $c=16$ case, so the situation may be more subtle.
The 24-dimensional case may be particularly interesting as the ground
state transforms trivially (as reviewed at the end of Section \ref{sec: CStheory}) under modular transformation of the torus.


\item It is possible to have an edge in which the interaction
varies along the edge so that $u_1$ is the only relevant operator
for $x<0$ and ${\tilde u}_1$ is the only relevant operator for
$x>0$. The edge will then be in the $E_8 \times E_8$ phase to the
left of the origin and the ${\rm Spin(32)}/\mathbb{Z}_2$ phase to the
right of the origin. It would be interesting to study the defect that
will be located at the origin.

\item Unimodular lattices occur in the study of four-manifold topology as the intersection form of $H^2(M, \mathbb{Z})$, where $M$ is a four-manifold and $H^2(M, \mathbb{Z})$ is the second cohomology group over the integers. 
(We assume that $M$ is closed.)
In the circumstances when de Rham cohomology can be defined, we can think of the intersection form as follows. 
Consider all pairs of 2-forms, $\omega_I, \omega_J$ and construct the matrix, $K_{IJ} = \int_M \omega_I \wedge \omega_J \in \mathbb{Z}$.
Even when de Rham cohomology does not make sense, the above matrix can be defined.
$K_{IJ}$ is unimodular and symmetric. 
Interestingly, the cases for which $K_{IJ}$ is even (and, therefore, provide intersection forms of the type studied in this paper) correspond to non-smooth four-manifolds.
The first instance is the so-called $E_8$ manifold whose intersection form is the $E_8$ Cartan matrix. 
Likewise, there exist two distinct four-manifolds, $E_8 \times E_8$ and the Chern manifold, with $E_8 \times E_8$ and ${\rm Spin}(32)/\mathbb{Z}_2$ intersection form, respectively.\cite{Freedman82} 
While these two four-manifolds are not equivalent or homeomorphic, they are cobordic: there exists a five-manifold whose two boundary components correspond to these two four-manifolds.
The cobordism can be understood as taking the direct sum of each four-manifold with $S^2 \times S^2$ which has intersection matrix equal to $U$.
A series of surgeries then relates these two connected augmented four-manifolds.
In other words, our paper has been a physical implementation of the above cobordism.
Is there a deeper connection between four-manifold topology and integer quantum Hall states?
We might go further and imagine that any such relation could be generalized to fractional and, possibly, non-abelian states.
Further, the introduction of symmetry-protected topological phases in 2+1d could inform the study of four-manifolds, i.e., the stabilizing symmetry of any phase could further refine the possible invariants characterizing any manifold.

\item We have concentrated on bosonic systems in this paper, but very similar considerations apply to fermionic SRE systems
with chiral edge modes, which correspond to positive-definite odd unimodular lattices.
The conventional integer quantum Hall states correspond to the hypercubic
lattices $\mathbb{Z}^N$. However, there is a second positive-definite
odd unimodular lattice in dimensions greater than $8$
namely $K_{E_8}\oplus I_{N-8}$. In dimensions greater than $11$,
there is also a third one, and there are still more in higher dimensions.
However, there is a unique unimodular lattice with
indefinite signature. Therefore, by a very similar construction
to the one that we have used here,
these different lattices correspond
to different edge phases of the $\nu\geq 9$ integer quantum Hall states.

\item Finally, stable equivalence is not restricted to topologically ordered states in 2+1d; it would be interesting to see explicitly how it manifests itself in the study of topological phases in other dimensions.

\end{itemize}

\acknowledgements
We would like to thank Parsa Bonderson, Matthew Fisher,
Michael Freedman, Tarun Grover, Max Metlitski, Ashvin Vishwanath,
and Jon Yard for discussions.
C.N. has been partially supported by the DARPA QuEST
program and AFOSR under grant FA9550-10-1- 0524.

\appendix

\section{Lattices and Matrices}
\label{sec:lattices-matrices}

In this appendix, we collect formulas for the various lattice vectors and matrices we use throughout the main text.

To fix some notation, consider the standard basis for ${\bf R}^N$,
\begin{eqnarray}
\label{vectdef}
{\hat{x}_I} = \begin{pmatrix} 0 \dotsb 0 & 1& 0 \dotsb 0 \end{pmatrix}^t, 
\end{eqnarray}
where the $1$ appears in the I-th row for $I = 1,..., N$.
The root lattice $\Gamma_G$ of any rank $N$ Lie group $G$ is defined in terms of linear combinations of the $\hat{x}_I$.
Given a basis ${\bf e}_I$ for the lattice, we may construct the Cartan matrix or $K$-matrix, $(K_G)_{IJ} =   e_I^a \eta_{a b} e_J^b$ where $\eta$ is the diagonal matrix  ${\rm diag}({\bf 1}^M, - {\bf 1}^{N - M})$ and ${\bf 1}^P$ is the $P$-component vector with every entry equal to unity.
The Cartan matrix summarizes the minimal data needed to specify a Lie group.
Geometrically, a diagonal entry $(K_G)_{II}$ is equal to the length-squared of the root $I$ and an off-diagonal entry $(K_G)_{IJ}$ gives the dot product between roots $I$ and $J$ and so can be interpreted as being proportional to the cosine of the angle (in ${\bf R}^N$) between the two roots.
Given the inverse $(K_G^{-1})^{IJ}$,
we may define dual lattice vectors $f^{I}_a = (K_G^{-1})^{IJ} \eta_{a b} e_J^b$
that satisfy $f^{I}_a e_J^a = \delta^I_J$.

\subsection{$\Gamma_{E_8}$}

A basis for the root lattice $\Gamma_{E_8}$ of the rank 8 group $E_8$ is given by
\begin{eqnarray}
\label{lieEeight}
{\bf e}_I & = & {\bf \hat{x}_I} - {\bf \hat{x}_{I + 1}},\ {\rm for}\ I = 1, ... 6,  \cr
{\bf e}_7 & = & - {\bf \hat{x}_1} - {\bf \hat{x}_2}, \cr
{\bf e}_8 & = & {1 \over 2} ({\bf \hat{x}_1} + ... + {\bf \hat{x}_8}).
\end{eqnarray}
The associated $K$-matrix takes the form,
\begin{eqnarray}
\label{Keight}
K_{E8} = \begin{pmatrix}
2 & -1 & 0 & 0 & 0 & 0 & 0 & 0 \cr
-1 & 2 & -1 & 0 & 0 & 0 & -1 & 0 \cr
0 & -1 & 2 & -1 & 0 & 0 & 0 & 0 \cr
0 & 0 & -1 & 2 & -1 & 0 & 0 & 0 \cr
0 & 0 & 0 & -1 & 2 & -1 & 0 & 0 \cr
0 & 0 & 0 & 0 & -1 & 2 & 0 & 0 \cr
0 & -1 & 0 & 0 & 0 & 0 & 2 & -1 \cr
0 & 0 & 0 & 0 & 0 & 0 & -1 & 2 \cr
\end{pmatrix}.
\end{eqnarray}
The inner product is Euclidean so $\eta_{a b} = \delta_{a b}$.

\subsection{$\Gamma_{E_8} \oplus \Gamma_{E_8}$}

The rank 16 Lie group $E_8 \times E_8$ is equal to two copies of $E_8$.
We take as our lattice basis for $\Gamma_{E_8} \oplus \Gamma_{E_8}$,
\begin{eqnarray}
\label{lieEeighttwo}
{\bf e}_I & = & {\bf \hat{x}_I} - {\bf \hat{x}_{I + 1}},\ {\rm for}\ I = 1, ... 6,  \cr
{\bf e}_7 & = & - {\bf \hat{x}_1} - {\bf \hat{x}_2}, \cr
{\bf e}_8 & = & {1 \over 2} ({\bf \hat{x}_1} + ... + {\bf \hat{x}_8}), \cr
{\bf e}_{8+ I} & = & {\bf \hat{x}_{9+I}} - {\bf \hat{x}_{10 +I}},\ {\rm for}\ I = 1, ..., 6, \cr
{\bf e}_{15} & = & {\bf \hat{x}_{15}} + {\bf \hat{x}_{16}}, \cr
{\bf e}_{16} & = & - {1 \over 2} ({\bf \hat{x}_9} + ... + {\bf \hat{x}_{16}}).
\end{eqnarray}
The associated $K$-matrix takes the form,
\begin{widetext}
\begin{eqnarray*}
\label{eqn:K-E8-E8}
K_{E_8 \oplus E_8}
  = \left(
\begin{array}{cccccccccccccccc}
2 & -1 & 0 & 0 & 0 & 0 & 0 & 0 & 0 & 0 & 0 & 0 & 0 & 0 & 0 & 0  \\
-1 & 2 & -1 & 0 & 0 & 0 & -1 & 0 & 0 & 0 & 0 & 0 & 0 & 0 & 0 & 0  \\
0 & -1 & 2 & -1 & 0 & 0 & 0 & 0 & 0 & 0 & 0 & 0 & 0 & 0 & 0 & 0 \\
0 & 0 & -1 & 2 & -1 & 0 & 0 & 0 & 0 & 0 & 0 & 0 & 0 & 0 & 0 & 0 \\
0 & 0 & 0 & -1 & 2 & -1 & 0 & 0 & 0 & 0 & 0 & 0 & 0 & 0 & 0 & 0  \\
0 & 0 & 0 & 0 & -1 & 2 & 0 & 0 & 0 & 0 & 0 & 0 & 0 & 0 & 0 & 0  \\
0 & -1 & 0 & 0 & 0 & 0 & 2 & -1 & 0 & 0 & 0 & 0 & 0 & 0 & 0 & 0\\
0 & 0 & 0 & 0 & 0 & 0 & -1 & 2 & 0 & 0 & 0 & 0 & 0 & 0 & 0 & 0 \\
0 & 0 & 0 & 0 & 0 & 0 & 0 & 0 & 2 & -1 & 0 & 0 & 0 & 0 & 0 & 0 \\
0 & 0 & 0 & 0 & 0 & 0 & 0 & 0 & -1 & 2 & -1 & 0 & 0 & 0 & 0 & 0 \\
0 & 0 & 0 & 0 & 0 & 0 & 0 & 0 & 0 & -1 & 2 & -1 & 0 & 0 & 0 & 0 \\
0 & 0 & 0 & 0 & 0 & 0 & 0 & 0 & 0 & 0 & -1 & 2 & -1 & 0 & 0 & 0 \\
0 & 0 & 0 & 0 & 0 & 0 & 0 & 0 & 0 & 0 & 0 & -1 & 2 & -1 & -1 & 0 \\
0 & 0 & 0 & 0 & 0 & 0 & 0 & 0 & 0 & 0 & 0 & 0 & -1 & 2 & 0 & 0 \\
0 & 0 & 0 & 0 & 0 & 0 & 0 & 0 & 0 & 0 & 0 & 0 & -1 & 0 & 2 & -1 \\
0 & 0 & 0 & 0 & 0 & 0 & 0 & 0 & 0 & 0 & 0 & 0 & 0 & 0 & -1 & 2 \\
\end{array}
\right).
\end{eqnarray*} 
\end{widetext}
The inner product is again taken to be $\eta_{ab} = \delta_{ab}$.

\subsection{$\Gamma_{{\rm Spin}(32)/Z_2}$}

A basis for the root lattice $\Gamma_{{\rm Spin}(32)/\mathbb{Z}_2}$  of the rank 16 Lie group ${\rm Spin}(32)/\mathbb{Z}_2$ is given by,
\begin{eqnarray}
\label{liespin}
\tilde{e}_I & = & {\bf \hat{x}_{I + 1}} - {\bf \hat{x}_{I + 2}},\ {\rm for}\ I = 1, ..., 14, \cr
\tilde{e}_{15} & = & {\bf \hat{x}_{15}} + {\bf \hat{x}_{16}}, \cr
\tilde{e}_{16} & = & - {1 \over 2} ({\bf \hat{x}_1} + ... + {\bf \hat{x}_{16}}).
\end{eqnarray}
The associated $K$-matrix,
\begin{widetext}
\begin{eqnarray*}
K_{{\rm Spin}(32)/\mathbb{Z}_2}
 = \left(
\begin{array}{cccccccccccccccccc}
2 & -1 & 0 & 0 & 0 & 0 & 0 & 0 & 0 & 0 & 0 & 0 & 0 & 0 & 0 & 0 \\
-1 & 2 & -1 & 0 & 0 & 0 & 0 & 0 & 0 & 0 & 0 & 0 & 0 & 0 & 0 & 0 \\
0 & -1 & 2 & -1 & 0 & 0 & 0 & 0 & 0 & 0 & 0 & 0 & 0 & 0 & 0 & 0 \\
0 & 0 & -1 & 2 & -1 & 0 & 0 & 0 & 0 & 0 & 0 & 0 & 0 & 0 & 0 & 0 \\
0 & 0 & 0 & -1 & 2 & -1 & 0 & 0 & 0 & 0 & 0 & 0 & 0 & 0 & 0 & 0 \\
0 & 0 & 0 & 0 & -1 & 2 & -1 & 0 & 0 & 0 & 0 & 0 & 0 & 0 & 0 & 0 \\
0 & 0 & 0 & 0 & 0 & -1 & 2 & -1 & 0 & 0 & 0 & 0 & 0 & 0 & 0 & 0 \\
0 & 0 & 0 & 0 & 0 & 0 & -1 & 2 & -1 & 0 & 0 & 0 & 0 & 0 & 0 & 0 \\
0 & 0 & 0 & 0 & 0 & 0 & 0 & -1 & 2 & -1 & 0 & 0 & 0 & 0 & 0 & 0 \\
0 & 0 & 0 & 0 & 0 & 0 & 0 & 0 & -1 & 2 & -1 & 0 & 0 & 0 & 0 & 0 \\
0 & 0 & 0 & 0 & 0 & 0 & 0 & 0 & 0 & -1 & 2 & -1 & 0 & 0 & 0 & 0 \\
0 & 0 & 0 & 0 & 0 & 0 & 0 & 0 & 0 & 0 & -1 & 2 & -1 & 0 & 0 & 0 \\
0 & 0 & 0 & 0 & 0 & 0 & 0 & 0 & 0 & 0 & 0 & -1 & 2 & -1 & -1 & 0 \\
0 & 0 & 0 & 0 & 0 & 0 & 0 & 0 & 0 & 0 & 0 & 0 & -1 & 2 & 0 & 0 \\
0 & 0 & 0 & 0 & 0 & 0 & 0 & 0 & 0 & 0 & 0 & 0 & -1 & 0 & 2 & -1 \\
0 & 0 & 0 & 0 & 0 & 0 & 0 & 0 & 0 & 0 & 0 & 0 & 0 & 0 & -1 & 4 \\
\end{array}
\right).
\end{eqnarray*}
\end{widetext}
The inner product is given by $\eta_{ab} = \delta_{ab}$.

\subsection{$\Gamma_{E_8} \oplus \Gamma_{E_8} \oplus U$}

To write a basis for the $\Gamma_{E_8} \oplus \Gamma_{E_8} \oplus U$ lattice, we must enlarge the dimension of our previous $\Gamma_{E_8} \oplus \Gamma_{E_8}$ lattice by two.
Thus, we take as our lattice basis,
\begin{eqnarray}
\label{lieEeighttwo+U}
{\bf e}_I & = & {\bf \hat{x}_I} - {\bf \hat{x}_{I + 1}},\ {\rm for}\ I = 1, ... 6,  \cr
{\bf e}_7 & = & - {\bf \hat{x}_1} - {\bf \hat{x}_2}, \cr
{\bf e}_8 & = & {1 \over 2} ({\bf \hat{x}_1} + ... + {\bf \hat{x}_8}), \cr
{\bf e}_{8+ I} & = & {\bf \hat{x}_{9+I}} - {\bf \hat{x}_{10 +I}},\ {\rm for}\ I = 1, ..., 6, \cr
{\bf e}_{15} & = & {\bf \hat{x}_{15}} + {\bf \hat{x}_{16}}, \cr
{\bf e}_{16} & = & - {1 \over 2} ({\bf \hat{x}_9} + ... + {\bf \hat{x}_{16}}), \cr
{\bf e}_{17} & = & \mbox{$\frac{1}{r}$} {\bf \hat{x}_{17}} + \mbox{$\frac{1}{r}$}{\bf \hat{x}_{18}}, \cr
{\bf e}_{18} & = & \,\,\mbox{$\frac{r}{2}$} {\bf \hat{x}_{17}} - \mbox{$\frac{r}{2}$} {\bf \hat{x}_{18}}.
\end{eqnarray}
The associated $K$-matrix takes the form,
\begin{widetext}
\begin{eqnarray*}
K_{E_8 \oplus E_8 \oplus U}
  = \left(
\begin{array}{cccccccccccccccccc}
2 & -1 & 0 & 0 & 0 & 0 & 0 & 0 & 0 & 0 & 0 & 0 & 0 & 0 & 0 & 0 & 0 & 0 \\
-1 & 2 & -1 & 0 & 0 & 0 & -1 & 0 & 0 & 0 & 0 & 0 & 0 & 0 & 0 & 0 & 0 & 0 \\
0 & -1 & 2 & -1 & 0 & 0 & 0 & 0 & 0 & 0 & 0 & 0 & 0 & 0 & 0 & 0 & 0 & 0\\
0 & 0 & -1 & 2 & -1 & 0 & 0 & 0 & 0 & 0 & 0 & 0 & 0 & 0 & 0 & 0 & 0 & 0\\
0 & 0 & 0 & -1 & 2 & -1 & 0 & 0 & 0 & 0 & 0 & 0 & 0 & 0 & 0 & 0 & 0 & 0 \\
0 & 0 & 0 & 0 & -1 & 2 & 0 & 0 & 0 & 0 & 0 & 0 & 0 & 0 & 0 & 0 & 0 & 0 \\
0 & -1 & 0 & 0 & 0 & 0 & 2 & -1 & 0 & 0 & 0 & 0 & 0 & 0 & 0 & 0 & 0 & 0\\
0 & 0 & 0 & 0 & 0 & 0 & -1 & 2 & 0 & 0 & 0 & 0 & 0 & 0 & 0 & 0 & 0 & 0\\
0 & 0 & 0 & 0 & 0 & 0 & 0 & 0 & 2 & -1 & 0 & 0 & 0 & 0 & 0 & 0 & 0 & 0\\
0 & 0 & 0 & 0 & 0 & 0 & 0 & 0 & -1 & 2 & -1 & 0 & 0 & 0 & 0 & 0 & 0 & 0\\
0 & 0 & 0 & 0 & 0 & 0 & 0 & 0 & 0 & -1 & 2 & -1 & 0 & 0 & 0 & 0 & 0 & 0\\
0 & 0 & 0 & 0 & 0 & 0 & 0 & 0 & 0 & 0 & -1 & 2 & -1 & 0 & 0 & 0 & 0 & 0\\
0 & 0 & 0 & 0 & 0 & 0 & 0 & 0 & 0 & 0 & 0 & -1 & 2 & -1 & -1 & 0 & 0 & 0\\
0 & 0 & 0 & 0 & 0 & 0 & 0 & 0 & 0 & 0 & 0 & 0 & -1 & 2 & 0 & 0 & 0 & 0\\
0 & 0 & 0 & 0 & 0 & 0 & 0 & 0 & 0 & 0 & 0 & 0 & -1 & 0 & 2 & -1& 0 & 0 \\
0 & 0 & 0 & 0 & 0 & 0 & 0 & 0 & 0 & 0 & 0 & 0 & 0 & 0 & -1 & 2 & 0 & 0\\
0 & 0 & 0 & 0 & 0 & 0 & 0 & 0 & 0 & 0 & 0 & 0 & 0 & 0 & 0 & 0 & 0 & 1\\
0 & 0 & 0 & 0 & 0 & 0 & 0 & 0 & 0 & 0 & 0 & 0 & 0 & 0 & 0 & 0 & 1 & 0\\
\end{array}
\right).
\end{eqnarray*} 
\end{widetext}
The inner product is taken with respect to $\eta_{ab} = ({\bf 1}^{17}, -1)$.

\subsection{$\Gamma_{{\rm Spin}(32)/\mathbb{Z}_2} \oplus U$}

We must again enlarge the dimension of $\Gamma_{{\rm Spin}(32)/\mathbb{Z}_2}$ by two in order to write a basis for $\Gamma_{{\rm Spin}(32)/\mathbb{Z}_2} \oplus U$,
\begin{eqnarray}
\label{liespin+U}
\tilde{e}_I & = & {\bf \hat{x}_{I + 1}} - {\bf \hat{x}_{I + 2}},\ {\rm for}\ I = 1, ..., 14, \cr
\tilde{e}_{15} & = & {\bf \hat{x}_{15}} + {\bf \hat{x}_{16}}, \cr
\tilde{e}_{16} & = & - {1 \over 2} ({\bf \hat{x}_1} + ... + {\bf \hat{x}_{16}}), \cr
\tilde{e}_{17} & = & - r {\bf \hat{x}_{17}} + r {\bf \hat{x}_{18}}, \cr
\tilde{e}_{18} & = & - \mbox{$\frac{1}{2r}$} {\bf \hat{x}_{17}} - \mbox{$\frac{1}{2r}$} {\bf \hat{x}_{18}}.
\end{eqnarray}
The associated $K$-matrix,
\begin{widetext}
\begin{eqnarray*}
K_{{\rm Spin}(32)/\mathbb{Z}_2 \oplus U}
 = \left(
\begin{array}{cccccccccccccccccc}
2 & -1 & 0 & 0 & 0 & 0 & 0 & 0 & 0 & 0 & 0 & 0 & 0 & 0 & 0 & 0 & 0 & 0  \\
-1 & 2 & -1 & 0 & 0 & 0 & 0 & 0 & 0 & 0 & 0 & 0 & 0 & 0 & 0 & 0 & 0 & 0 \\
0 & -1 & 2 & -1 & 0 & 0 & 0 & 0 & 0 & 0 & 0 & 0 & 0 & 0 & 0 & 0 & 0 & 0 \\
0 & 0 & -1 & 2 & -1 & 0 & 0 & 0 & 0 & 0 & 0 & 0 & 0 & 0 & 0 & 0 & 0 & 0 \\
0 & 0 & 0 & -1 & 2 & -1 & 0 & 0 & 0 & 0 & 0 & 0 & 0 & 0 & 0 & 0 & 0 & 0 \\
0 & 0 & 0 & 0 & -1 & 2 & -1 & 0 & 0 & 0 & 0 & 0 & 0 & 0 & 0 & 0 & 0 & 0 \\
0 & 0 & 0 & 0 & 0 & -1 & 2 & -1 & 0 & 0 & 0 & 0 & 0 & 0 & 0 & 0 & 0 & 0 \\
0 & 0 & 0 & 0 & 0 & 0 & -1 & 2 & -1 & 0 & 0 & 0 & 0 & 0 & 0 & 0 & 0 & 0 \\
0 & 0 & 0 & 0 & 0 & 0 & 0 & -1 & 2 & -1 & 0 & 0 & 0 & 0 & 0 & 0 & 0 & 0 \\
0 & 0 & 0 & 0 & 0 & 0 & 0 & 0 & -1 & 2 & -1 & 0 & 0 & 0 & 0 & 0 & 0 & 0 \\
0 & 0 & 0 & 0 & 0 & 0 & 0 & 0 & 0 & -1 & 2 & -1 & 0 & 0 & 0 & 0 & 0 & 0 \\
0 & 0 & 0 & 0 & 0 & 0 & 0 & 0 & 0 & 0 & -1 & 2 & -1 & 0 & 0 & 0 & 0 & 0 \\
0 & 0 & 0 & 0 & 0 & 0 & 0 & 0 & 0 & 0 & 0 & -1 & 2 & -1 & -1 & 0 & 0 & 0 \\
0 & 0 & 0 & 0 & 0 & 0 & 0 & 0 & 0 & 0 & 0 & 0 & -1 & 2 & 0 & 0 & 0 & 0 \\
0 & 0 & 0 & 0 & 0 & 0 & 0 & 0 & 0 & 0 & 0 & 0 & -1 & 0 & 2 & -1 & 0 & 0 \\
0 & 0 & 0 & 0 & 0 & 0 & 0 & 0 & 0 & 0 & 0 & 0 & 0 & 0 & -1 & 4 & 0 & 0 \\
0 & 0 & 0 & 0 & 0 & 0 & 0 & 0 & 0 & 0 & 0 & 0 & 0 & 0 & 0 & 0 & 0 & 1 \\
0 & 0 & 0 & 0 & 0 & 0 & 0 & 0 & 0 & 0 & 0 & 0 & 0 & 0 & 0 & 0 & 1 & 0 \\
\end{array}
\right).
\end{eqnarray*}
\end{widetext}
The inner product is taken with respect to $\eta_{ab} = ({\bf 1}^{17}, -1)$.

\subsection{$SO(17,1)$ and $SL(18, \mathbb{Z})$ Transformations}

There exist two distinct even, self-dual 16-dimensional lattices, $\Gamma_{E_8} \oplus \Gamma_{E_8}$ and $\Gamma_{{\rm Spin}(32)/\mathbb{Z}_2}$, that cannot be rotated into each other via an $SO(16)$ transformation \cite{Serre73}.
However, if we augment each lattice by $U$, we obtain a Lorentzian lattice of signature $(17,1)$, i.e., the augmented lattice has the inner product $\eta_{ab} = {\rm diag}({\bf 1}^{17}, -1)$.
Such lattices are unique up to an $SO(17,1)$ rotation.
Following \cite{Ginsparg87}, the $SO(17,1)$ transformation relating the $\Gamma_{E_8} \oplus \Gamma_{E_8} \oplus U$ and $\Gamma_{{\rm Spin}(32)/\mathbb{Z}_2} \oplus U$ lattices is given by,
\begin{widetext}
\begin{eqnarray*}
O_G  = \left(
\begin{array}{cccccccccccccccccc}
 1 & 0 & 0 & 0 & 0 & 0 & 0 & -1 & 1 & 0 & 0 & 0 & 0 & 0 & 0 & 0 & \frac{1}{2 r}-\frac{-1+r^2}{2 r} & -\frac{1}{2 r}-\frac{1+r^2}{2 r} \\
 0 & 1 & 0 & 0 & 0 & 0 & 0 & -1 & 1 & 0 & 0 & 0 & 0 & 0 & 0 & 0 & \frac{1}{2 r}-\frac{-1+r^2}{2 r} & -\frac{1}{2 r}-\frac{1+r^2}{2 r} \\
 0 & 0 & 1 & 0 & 0 & 0 & 0 & -1 & 1 & 0 & 0 & 0 & 0 & 0 & 0 & 0 & \frac{1}{2 r}-\frac{-1+r^2}{2 r} & -\frac{1}{2 r}-\frac{1+r^2}{2 r} \\
 0 & 0 & 0 & 1 & 0 & 0 & 0 & -1 & 1 & 0 & 0 & 0 & 0 & 0 & 0 & 0 & \frac{1}{2 r}-\frac{-1+r^2}{2 r} & -\frac{1}{2 r}-\frac{1+r^2}{2 r} \\
 0 & 0 & 0 & 0 & 1 & 0 & 0 & -1 & 1 & 0 & 0 & 0 & 0 & 0 & 0 & 0 & \frac{1}{2 r}-\frac{-1+r^2}{2 r} & -\frac{1}{2 r}-\frac{1+r^2}{2 r} \\
 0 & 0 & 0 & 0 & 0 & 1 & 0 & -1 & 1 & 0 & 0 & 0 & 0 & 0 & 0 & 0 & \frac{1}{2 r}-\frac{-1+r^2}{2 r} & -\frac{1}{2 r}-\frac{1+r^2}{2 r} \\
 0 & 0 & 0 & 0 & 0 & 0 & 1 & -1 & 1 & 0 & 0 & 0 & 0 & 0 & 0 & 0 & \frac{1}{2 r}-\frac{-1+r^2}{2 r} & -\frac{1}{2 r}-\frac{1+r^2}{2 r} \\
 0 & 0 & 0 & 0 & 0 & 0 & 0 & 0 & 1 & 0 & 0 & 0 & 0 & 0 & 0 & 0 & -\frac{1}{2 r}-\frac{-1+r^2}{2 r} & \frac{1}{2 r}-\frac{1+r^2}{2 r} \\
 0 & 0 & 0 & 0 & 0 & 0 & 0 & 0 & 1 & 0 & 0 & 0 & 0 & 0 & 0 & 0 & \frac{1}{r} & -\frac{1}{r} \\
 0 & 0 & 0 & 0 & 0 & 0 & 0 & 0 & 0 & 1 & 0 & 0 & 0 & 0 & 0 & 0 & 0 & 0 \\
 0 & 0 & 0 & 0 & 0 & 0 & 0 & 0 & 0 & 0 & 1 & 0 & 0 & 0 & 0 & 0 & 0 & 0 \\
 0 & 0 & 0 & 0 & 0 & 0 & 0 & 0 & 0 & 0 & 0 & 1 & 0 & 0 & 0 & 0 & 0 & 0 \\
 0 & 0 & 0 & 0 & 0 & 0 & 0 & 0 & 0 & 0 & 0 & 0 & 1 & 0 & 0 & 0 & 0 & 0 \\
 0 & 0 & 0 & 0 & 0 & 0 & 0 & 0 & 0 & 0 & 0 & 0 & 0 & 1 & 0 & 0 & 0 & 0 \\
 0 & 0 & 0 & 0 & 0 & 0 & 0 & 0 & 0 & 0 & 0 & 0 & 0 & 0 & 1 & 0 & 0 & 0 \\
 0 & 0 & 0 & 0 & 0 & 0 & 0 & 0 & 0 & 0 & 0 & 0 & 0 & 0 & 0 & 1 & 0 & 0 \\
 \frac{r}{2} & \frac{r}{2} & \frac{r}{2} & \frac{r}{2} & \frac{r}{2} & \frac{r}{2} & \frac{r}{2} & -\frac{r}{2}+\frac{1-r^2}{r} & r-\frac{1-r^2}{r} & 0 & 0 & 0 & 0 & 0 & 0 & 0 & \frac{1}{2}+\frac{\left(1-r^2\right) \left(-1+r^2\right)}{r^2} & -\frac{1}{2}-r^2+\frac{1-r^2}{r^2} \\
 -\frac{r}{2} & -\frac{r}{2} & -\frac{r}{2} & -\frac{r}{2} & -\frac{r}{2} & -\frac{r}{2} & -\frac{r}{2} & \frac{r}{2}+\frac{1+r^2}{r} & -r-\frac{1+r^2}{r} & 0 & 0 & 0 & 0 & 0 & 0 & 0 & -\frac{1}{2}+r^2-\frac{1+r^2}{r^2} & \frac{1}{2}+\frac{\left(1+r^2\right)^2}{r^2} \\
\end{array}
\right).
\end{eqnarray*}
\end{widetext}
$O_G$ acts on basis vectors as 
\begin{eqnarray}
\label{basisrotation}
O^a_{G\ b} {e}_I^b = \sum_J m_I^J \tilde{e}_J^a,
\end{eqnarray}
where $m_I^J$ are a collection of integers.

Because $O_G$ lies in the component of $SO(17,1)$ connected to the identity transformation, we may build $O_G$ from a series of infinitesimal transformations beginning at ${\bf 1}$.
First, we rewrite, 
\begin{eqnarray}
\label{expon}
O_G = \eta W(A) \eta W(A'),
\end{eqnarray}
where
\begin{align}
W(A) & = \exp\left[ \frac{1}{2} 
		\left( 
		\begin{array}{ccc}
			0 & A & -A \\
			-A^t & 0 & 0 \\
			-A^t & 0 &  0
		\end{array}
		\right) \right], \quad {\rm with} \\
A & = \frac{2}{r} \left( 0^7, -1, 1, 0^7 \right), \\
A' & = -2r \left( \left(\frac{1}{2}\right)^8, 0^8 \right).
\end{align}
We then introduce the (infinitesimal) parameter $s$ by rescaling $A, A' \rightarrow s A, s A'$ and defining,
\begin{eqnarray}
\label{infinites}
O_G(s) = \eta W(s A) \eta W(s A').
\end{eqnarray}
(While the resulting matrix does not fit between the margins of this page, the expression is not beautiful.)

Substituting the transformation Eq. \eqref{basisrotation} into the periodicity condition, $X^a \equiv X^a + 2 \pi n^I e_I^a$, for the $\Gamma_{E_8} \oplus \Gamma_{E_8} \oplus U$ lattice, we find: 
\begin{eqnarray}
\label{transform}
(O_G)^a_{\ b} X^b \equiv (O_G)^a_{\ b} X^b + 2 \pi \tilde{n}^J \tilde{e}_J^a,
\end{eqnarray}
where we have defined the integer vector $\tilde{n}^J = \sum_I n^I m_I^J$.
However, Eq. \eqref{transform} is simply the periodicity obeyed by $\tilde{X}^a$. 
Therefore, we identify $\tilde{X}^a = (O_G)^a_{\ b} X^b$.
Having identified $X^a$ and $\tilde{X^b}$ through the $SO(17,1)$ transformation $O_G$, we can obtain the $SL(18,{\mathbb{Z}})$ transformation $W_G$ that relates $K_{{\rm Spin}(32)/\mathbb{Z}_2 \oplus U}$ and  $K_{E_8 \oplus E_8 \oplus U}$ by conjugation.
The desired transformation is read off from the relation,
\begin{eqnarray}
\label{linfracbig2}
\tilde{\phi}^J = \tilde{f}^{J}_a (O_G)^a_{\ b} e_I^b \phi^I =: (W_G)_{I J} \phi^I,
\end{eqnarray}
which follows immediately from Eq. (\ref{basisrotation}).
We find:
\begin{widetext}
\begin{eqnarray*}
W_G = \left(
\begin{array}{cccccccccccccccccc}
-2 & 1 & 0 & 0 & 0 & 0 & 0 & 0 & 0 & 0 & 0 & 0 & 0 & 0 & 0 & 0 & 0 & 0  \\
-3 & 0 & 1 & 0 & 0 & 0 & 1 & 0 & 0 & 0 & 0 & 0 & 0 & 0 & 0 & 0 & 0 & 0 \\
-4 & 0 &0 & 1 & 0 & 0 & 2 & 0 & 0 & 0 & 0 & 0 & 0 & 0 & 0 & 0 & 0 & 0 \\
-5 & 0 & 0& 0 & 1 & 0 & 3 & 0 & 0 & 0 & 0 & 0 & 0 & 0 & 0 & 0 & 0 & 0 \\
-6 & 0 & 0 & 0 & 0 & 1 & 4 & 0 & 0 & 0 & 0 & 0 & 0 & 0 & 0 & 0 & 0 & 0 \\
-7 & 0 & 0 & 0 & 0 & 0 & 5& 0 & 0 & 0 & 0 & 0 & 0 & 0 & 0 & 0 & 0 & 0 \\
-8 & 0 & 0 & 0 & 0 & 0 & 6 & 0 & 0 & 0 & 0 & 0 & 0 & 0 & 0 & 0 & 0 & -1 \\
-9 & 0 & 0 & 0 & 0 & 0 & 7& 0 & 0 & 0 & 0 & 0 & 0 & 0 & 0 & 0 & 1 & -1 \\
-10 & 0 & 0 & 0 & 0 & 0 & 8 & 0 & 1 & 0 & 0 & 0 & 0 & 0 & 0 & 0 & 2 & -2 \\
-11 & 0 & 0 & 0 & 0 & 0 & 9 & 0 & 0 & 1 & 0 & 0 & 0 & 0 & 0 & 0 & 3 & -3 \\
-12 & 0 & 0 & 0 & 0 & 0 & 10 & 0 & 0 & 0 & 1 & 0 & 0 & 0 & 0 & 0 & 4 & -4 \\
-13 & 0 & 0 & 0 & 0 & 0 & 11 & 0 & 0 & 0 & 0 & 0 & 1 & 0 & 0 & 0 & 5 & -5 \\
-14 & 0 & 0 & 0 & 0 & 0 & 12 & 0 & 0 & 0 & 0 & 0 & 1 & 0 & 0 & 0 & 6 & -6 \\
-7 & 0 & 0 & 0 & 0 & 0 & 6 & 0 & 0 & 0 & 0 & 0 & 0 & 1 & 0 & 0 & 3 & -3 \\
-8 & 0 & 0 & 0 & 0 & 0 & 7 & 0 & 0 & 0 & 0 & 0 & 0 & 0 & 1 & 0 & 4 & -4 \\
-2 & 0 & 0 & 0 & 0 & 0 & 2 & 0 & 0 & 0 & 0 & 0 & 0 & 0 & 0& 1 & 2 & -2 \\
0 & 0 & 0 & 0 & 0 & 0 & 1 & -1 & 0 & 0 & 0 & 0 & 0 & 0 & 0 & 1 & 2 & -2 \\
0 & 0 & 0& 0 & 0 & 0 & 0 & -1 & 0 & 0 & 0 & 0 & 0 & 0 & 0 & -1 & -2 & 2 \\
\end{array}
\right).
\end{eqnarray*}
\end{widetext}
This matrix satisfies $W^T_G K_{{\rm Spin}(32)/\mathbb{Z}_2 \oplus U} W = K_{E_8 \oplus E_8 \oplus U}$.

\section{``Dimension Contraction'' and Relevant Mass-Generating
Operators at Intermediate $r, s$}
\label{sec:dimension-contraction}

We consider spin-0 operators that take the form
$\cos( p_a X^a)$, with ${p_a}\in {\Gamma_8}\oplus{\Gamma_8}\oplus U$
and $\eta^{ab}{p_a}{p_b}=0$. Even if
$\frac{1}{2}\delta^{ab}{p_a}{p_b}>2$, which means that
$\cos( p_a X^a)$ is irrelevant at $s=0$, this operator may become relevant
at an intermediate value of $s$.
At general $s$, the scaling dimension of the operator is  $\frac{1}{2} \delta^{ab}{q_a}{q_b}=|q_{18}|^2$, where $q_b = p_a (O^{-1}_G(s))^a_{\ b}$. 
In writing the scaling dimension in terms of $q_{18}$ only, we have used the fact that $q_b$ is a null vector in $\mathbb{R}^{17,1}$ ($\eta^{a b} q_a q_b = q_1^2 + ... + q_{17}^2 - q_{18}^2 = 0$).
Thus, $\cos(p_a X^a)$ will become relevant at $s$ if $p_a (O_G^{-1}(s))^a_{18}$ is sufficiently Lorentz contracted so that $q_{18}^2 < 2$.

If the direction of the boost $O_G^{-1}(s)$ happened to be along the 1-direction, then we know that the only components of $p_a$ affected by the boost are the 1st and 18th component;
they are contracted/dilated according to:
\begin{equation}
\begin{pmatrix}
p_1 \cr p_{18} 
\end{pmatrix}
\mapsto
\begin{pmatrix}
\cosh(\alpha) & - \sinh(\alpha) \cr - \sinh(\alpha) & \cosh(\alpha)
\end{pmatrix}
\begin{pmatrix}
p_1 \cr p_{18}
\end{pmatrix}.
\end{equation}
Therefore, multiples of the eigenvectors $(1, \pm 1)^T$ with
eigenvalues $\exp(\mp \alpha)$ have components that are maximally contracted/dilated.
If the boost took the above simple form, it would be simple to choose a vector $p_a$ whose 18th component after the boost was maximally contracted.
This vector would determine the most relevant operator at a given point in the $(r, s)$ phase diagram.

Unfortunately, $O_G^{-1}(s)$ is defined in terms of a rather complicated combination of rotations and boosts, and so it is not a priori obvious which spatial direction to choose in order to maximize the possible contraction, i.e., it is difficult to know the direction $\vec{v}$ of the boost. 
However, we know that we can view the $O_G^{-1}(s)$ transformation as: $O_G^{-1}(s) = M^{T} \Lambda M$, where $M$ is a rotation that aligns $\vec{v}$ along the 1-direction and $\Lambda$ is a boost along the 1-direction.
(Both of these transformations, of course, depend upon the initially chosen $r$ and $s$.)
To find null vectors whose components maximally contract, we need only consider the eigenvector of $O_G^{-1}(s)$ given by $M^{{\rm tr}} (1, 0^{16}, 1)^{{\rm tr}}$ with eigenvalue $\exp(- \alpha)$, for some constant $\alpha$ depending upon $r$ and $s$.
For $(r, s) = (3, 3/5)$ we find that this maximally contracting eigenvector takes the simple (approximate) form:
\begin{equation}
p_a = .3 f^7_{\ a} + (.1 - .6) f^8_{\ a} + .1 f^{16}_{\ a} + f^{17}_{\ a} - .9 f^{18}_{\ a}.    
\end{equation}
While the components of this vector are maximally contracted under $O_G^{-1}(s)$ in the sense discussed above, it is certainly not an element of $\Gamma_{E_8} \oplus \Gamma_{E_8} \oplus U$ since the coefficients are not integral.
We can find a vector with very large components that is nearly parallel to this vector,
but it will be irrelevant because $O_G^{-1}(s)$ cannot contract it by enough at
$(r, s) = (3, 3/5)$.

However, we can find a shorter lattice vector that is sufficiently aligned with the maximally contracting vector, but of lower starting dimension so that we obtain
a relevant operator at the point of interest.
Indeed, if we take the ansatz:
\begin{equation}
p_a = n f^7_{\ a} + (m - 2 n) f^8_{\ a} + m f^{16}_{\ a} + n_{17} f^{17}_{\ a} + n_{18} f^{18}_{\ a},
\end{equation}
it is straightforward to find $n, m, n_{17}$ and $n_{18}$ determining a relevant spin-0 operator at $(r,s)$. At $(r, s) = (3, 3/5)$,
we may take $n = 1, m = 2, n_{17} = 2$ and $n_{18} = -3$.
We lack a proof that this ansatz is sufficient to exclude
all possible non-chiral points in the $(r,s)$ phase diagram. 
However, we have yet to find a point $(r,s)$ for which this ansatz
is unsuccessful. Thus, we expect the non-chiral phase to be entirely
removed by this collection of operators combined with those discussed earlier.
(Note, we expect the resulting chiral phase for this operator to be
${\rm Spin}(32)/\mathbb{Z}_2$.)

\bibliography{heterotic}


\end{document}